\documentclass[12pt]{article}
\usepackage[centertags]{amsmath}
\usepackage{setspace}
\usepackage{amsfonts}
\usepackage{amssymb}
\usepackage{amsthm}
\newcommand{\beq}{\begin{equation}}
\newcommand{\eeq}[1]{\label{#1}\end{equation}}
\newcommand{\bea}{\begin{eqnarray}}
\newcommand{\eea}[1]{\label{#1}\end{eqnarray}}
\begin{document}
\setlength{\topmargin}{-1cm} \setlength{\oddsidemargin}{0cm}
\setlength{\evensidemargin}{0cm}
\begin{titlepage}
%\hfill  {\small NYU-TH/07-137}
\begin{center}
{\Large \bf A Model Independent Ultraviolet Cutoff for Theories with
Charged Massive Higher Spin Fields}

\vspace{20pt}

{\large Massimo Porrati$^{a,b}$ and Rakibur Rahman$^a$}

\vspace{12pt}
$a$) Center for Cosmology and Particle Physics\\
     Department of Physics, New York University\\
     4 Washington Place, New York, NY 10003\\
\vspace{6pt}
$b$) Department of Physics, Columbia University\\
     New York, NY 10027\\

\end{center}

\vspace{20pt}

\begin{abstract}

We argue that the theory of a massive higher spin field coupled to
electromagnetism in flat space possesses an intrinsic, model
independent, finite upper bound on its UV cutoff. By employing the
St\"{u}ckelberg formalism we do a systematic study to quantify the
degree of singularity of the massless limit in the cases of spin 2,
3, 3/2, and 5/2. We then generalize the results for arbitrary spin
to find an expression for the maximum cutoff of the theory as a
function of the particle's mass, spin, and electric charge. We also
briefly explain the physical implications of the result and discuss
how it could be sharpened by use of causality constraints.

\end{abstract}

\end{titlepage}

\newpage

\section{Introduction}
Powerful arguments exist that forbid massless particles from
interacting with electromagnetism (EM) or gravity when their spin
$s$ exceeds a certain maximum value~\cite{ad,ww,p}. This is $s=1$
for EM, and $s=2$ for gravity. Charged, massive particles instead do
exist. Hadronic resonances and open-string charged states are but
two examples. Even more obviously, any classical charged, spinning
object can be decomposed into irreducible representations of the
Poincar\'e group, and is thus mathematically the same as a particle.
If common sense rebels against treating a charged macroscopic top in
the same manner as we treat an electron, it is because we usually
(and correctly) associate other properties to the objects we call
{\em elementary} particles. Chief among them is that they interact
as point-like objects up to distance scales parametrically larger
than their Compton wavelength. A classical spinning top clearly does
not satisfy this condition. Neither do high-spin hadronic
resonances. They are quark bound states whose mass is always much
larger than their inverse size: the former is well above $1.5\,
GeV$, while the latter is $\mathcal{O}(\Lambda_{QCD})\approx 400\,
MeV$. Even seemingly true elementary particles as charged high-spin
excitations of the open sting are not truly point-like. Their mass
is always larger than the sting scale $M_S$, i.e. the same parameter
that also sets the intrinsic non-locality scale of string theory.

Are these properties accidents? Can a high-spin massive particle be
described by a local Lagrangian up to arbitrarily high energy
scales? The answer to this question is no, because the no-go
theorems on EM and gravitationally coupled massless particles cited
above imply that the cutoff of that Lagrangian must vanish in the
massless limit $m\rightarrow 0$.

To find the explicit parametric dependence of the cutoff on the mass
and the relevant coupling constant is the much more difficult
problem we study in this paper. This task requires additional
assumptions about the interacting high-spin particle; of course, the
more the assumptions, the stronger the result. We aim at obtaining
model-independent, universal limits on the cutoff of the effective
action describing high-spin particles. For this reason we will not
get the strongest possible bound on the cutoff $\Lambda$, but rather
an upper bound that no theory can beat. Let us first state the limit
we will obtain, and later discuss its meaning and the assumptions we
need for its derivation.

The largest cutoff of the local effective action describing a
massive charged particle of spin larger than one, coupled to EM (or,
more generally to a massless, Abelian vector) is
\beq
\Lambda_s \leq
C me^{-1/(2s-1)}, \qquad C=\mathcal{O}(1)~ \text{constant}.
\eeq{m1}
This formula is valid in the limit $e\ll 1$. With $e\approx 0.3$, EM
coupling is at the limit of the range of validity of Eq.~(\ref{m1}).
Still, Eq.~(\ref{m1}) is not empty: it states that high-spin
particles admit a local description even for energies above their
mass, when they must be treated as true dynamical degrees of freedom
in the effective action.

As we already mentioned, the cutoff~(\ref{m1}) is an upper bound. In
fact, the true cutoff is lower in all known examples of UV complete
theories containing high-spin, charged massive particles. The most
physical example of this fact is QCD. The state of affairs is even
clearer in open string theory, where one finds an infinite tower of
massive charged particles with spin larger than one. The dynamics of
any one of those particles in an external {\em constant} EM field
can be studied independently of all others. It is described by a
non-minimal yet local Lagrangian due to Argyres and Nappi~\cite{an}.
The Lagrangian propagates five degrees of freedom inside the
standard Lorentzian light cone~\cite{an}. It is thus exempt from the
pathologies found years ago by Velo and Zwanziger~\cite{vz} in their
study of EM-coupled massive spin-3/2 and 2. The Argyres-Nappi
Lagrangian~\cite{an} escapes those problems at the cost of being
extremely non-minimal: it contains an infinite tower of
``quadrupole'' non-renormalizable couplings involving two
derivatives of the spin-2 field $h_{\mu\nu}$ and arbitrary powers of
the EM field strength $F_{\mu\nu}$. Somewhat symbolically their
generic form is \beq h^*_{\mu\nu} [(\alpha' e F)^n
\partial^2]^{\mu\nu}_{~~\rho\sigma}h^{\rho\sigma}. \eeq{m2}
The cutoff scale at which a Lagrangian containing non-renormalizable
terms such as~(\ref{m2}) breaks down is $\alpha'^{-1/2} = M_S$.
Validity of the open-string perturbation series requires a string
coupling constant $g_S = e^2 \ll 1$; thus, the cutoff of the
effective Argyres-Nappi action is much smaller than the ``optimal''
one.

The case of charged, massive spin-1 is worth special attention,
because some of the technical procedures we employ in this
paper do not work there, for reasons we shall explain later. Nevertheless,
Eq.~(\ref{m1}) still holds, since the cutoff of a Lagrangian
containing {\em only} a charged massive spin-1 and the EM field is
indeed ${\cal O}(m/e)$. In this case, we also know explicit UV completions of
such a theory. One such completion is to embed the spin-1 field as a
vector of an $SU(2)$ gauge group broken to $U(1)$ by a Higgs field,
which is also a vector of $SU(2)$. Then, the bound~(\ref{m1})
simply means that either an extra degree of freedom (the neutral
Higgs) exists, with mass smaller than $\Lambda_1=m/e$, or the theory
becomes strongly interacting and unitarizes at $\Lambda_1$.

To summarize, the cutoff~(\ref{m1}) is model independent, but its
very generality means that we have very little information about the
UV completion at or above that energy scale. To understand the
technique we will be using to derive Eq.~(\ref{m1}), let us recall
what we are looking for. First of all, we shall limit ourselves to
finding an effective action that generates sensible low-energy
scattering amplitudes between states with a finite number of hard
quanta. So, we will not attempt to solve the Velo-Zwanziger
causality problem or any other pathology that may arise in {\em
external} non-trivial backgrounds. To compute these S-matrix
elements, we need an effective {\em local} Lagrangian for a charged
massive particle of arbitrary spin. While many choices exist in the
literature for free Lagrangians, our choice is one with as few
auxiliary fields as possible,  e.g. the Singh-Hagen
Lagrangian~\cite{sh}. All those Lagrangians enjoy restricted gauge
invariances in the massless limit. These gauge invariances are
broken by the interactions. This phenomenon is at the root of the
no-go theorems on interacting massless particles~\cite{ad,ww,p}. It
also makes it hard to find the UV cutoff of the interacting
Lagrangians, since gauge invariances of the kinetic term change the
canonical dimensions of the fields. The way out of this problem is
well known: introduce ``compensating'' degrees of freedom
(St\"uckelberg fields) which restore the gauge invariance, and fix
this newly introduced gauge invariance to set all kinetic terms to a
canonical form $-$ i.e. a form that gives canonical dimensions to
all fields. Once all fields have canonical dimensions (1 for Bosons
and $3/2$ for Fermions), then the interacting Lagrangian will
acquire certain non-renormalizable interactions involving the
St\"uckelberg fields. These will be weighted by coupling constants
with negative mass dimensions; symbolically \beq L=
L_{\text{renormalizable}} + \sum_{n>0} (\Lambda_n)^{-n} O_{n+4},
\eeq{m3} where $O_{n+4}$ denotes operators of dimension ($n+4$).
Some of these operators can be eliminated by field redefinitions or
by adding non-minimal terms to the Lagrangian $L$; the smallest
$\Lambda_n$ in the surviving terms defines the ultimate cutoff of
our effective Lagrangian. The procedure to follow can be described
more fully and more technically in a few steps, namely:
\begin{enumerate}
\item
Write a (non-gauge invariant) massive Lagrangian
with minimal number of auxiliary fields (e.g., {\em \`a la} Singh
and Hagen).
\item
Introduce St\"uckelberg fields and St\"uckelberg
gauge symmetry. Any auxiliary field appearing in the Lagrangian in
step 1 that is not a (gamma)trace of the high-spin field must be
identified as a trace of a St\"uckelberg field by appropriate field
redefinitions. For such a field one obtains a gauge invariance for free.
\item
Complexify the fields, if required, and introduce interaction with a
new gauge field (e.g., electromagnetism) by replacing ordinary
derivatives with covariant ones.
\item
Diagonalize all kinetic terms, i.e., get rid of kinetic
mixing by field redefinitions and/or by covariant gauge fixing terms.
\item
Look for the most divergent term(s) in the
Lagrangian, in an appropriate limit of zero mass and zero coupling.
These terms will involve fields that are zero (i.e. gauge) modes of
the free kinetic operator before gauge fixing. One needs to take
care of the non-commutativity of covariant derivatives and correctly
interpret terms proportional to the equations of motion.
\item
Try to remove non-renormalizable terms by adding
non-minimal terms. This may not always be possible.
\item
Find the cutoff of the effective field theory, and
interpret the physics implied by the divergent term(s).
\end{enumerate}

Steps 1 and 2 lead us to a gauge invariant description of massive
high spin fields $-$ the so-called St\"uckelberg
formalism~\cite{z1,z2,met,med,bian,hall}\footnote{There exist other
kind of gauge invariant descriptions for massive higher spins, e.g.
the BRST method~\cite{brst}, the frame-like
formulation~\cite{frame}, and the quartet formulation~\cite{bg}.}.
Such gauge invariant description is convenient in that it allows one
to introduce interactions simply by replacing ordinary derivatives
with covariant ones (it is also useful for studying partially
massless theories that appear in (A)dS space-time~\cite{partial}).
One can obtain the St\"{u}ckelberg invariant Lagrangian by starting
with the massless Lagrangian~\cite{ff} in (4+1)D, and then
Kaluza-Klein reducing it to (3+1)D~\cite{ady,rs}. The higher
dimensional gauge invariance gives rise to the St\"{u}ckelberg
symmetry in lower dimension. By suitably gauge fixing the resulting
Lagrangian, one gets a (non-gauge invariant) massive Lagrangian with
minimal number of auxiliary fields, as mentioned in Step 1, which is
equivalent to the Singh-Hagen Lagrangian~\cite{sh} up to field
redefinitions. Then the St\"uckelberg trick in Step 2 is done with
the help of carefully constructed gauge invariant tensors.

The procedure outlined above was carried out for spin-2 fields
in~\cite{pr1,pr2}; in Section 2, we review the main points of these
references and give an improved derivation of the cutoff
$\Lambda_2=me^{-1/3}$. Section 3 is devoted to studying EM
interactions of a spin-3 particle, where for the first time we find
a new complication, namely an extra auxiliary scalar field that
cannot be set to zero by gauge transformations. Sections 4 and 5
apply the St\"uckelberg method to charged Fermions; first spin-3/2
Fermions, then spin-5/2. Section 6 generalizes the findings of all
previous Sections to arbitrary spin. There the bound~(\ref{m1}) is
at last derived. Section 7 summarizes our findings, and briefly
discusses a few additional topics related to the physics of
interacting high spin fields. In particular, taking heed of the well
known case of charged spin-1 particles, we will describe there some
alternative possibilities for the UV completion of a theory of
higher spin particles. We will also suggest that causality
constraints in external background fields may give stronger model
independent bounds on the UV cutoff.

\section{Massive Spin-2 Field Coupled to EM}

The electromagnetic interaction of massive spin-2 field has been studied
by various authors~\cite{vz,z2,pr1,pr2,fksc,z3,z4,d2}. Here we consider flat
space-time background.

First we write down the Pauli-Fierz Lagrangian~\cite{pfn} with St\"{u}ckelberg
fields, complexify the fields, and then replace ordinary derivatives
with covariant ones. Thus we obtain \beq L= -\,|D_\mu
\tilde{h}_{\nu\rho}|^2+2|D_\mu \tilde{h}^{\mu\nu}|^2 +|D_\mu
\tilde{h}|^2 - [D_\mu \tilde{h}^{*\mu\nu}D_\nu \tilde{h} +
\text{c.c.}] -
m^2[\tilde{h}_{\mu\nu}^*\tilde{h}^{\mu\nu}-\tilde{h}^*\tilde{h}],
\eeq{r7} with \beq \tilde{h}_{\mu\nu} = h_{\mu\nu} +
\frac{1}{m}\,D_{\mu}\left(B_\nu-\frac{1}{2m}D_{\nu}\phi\right) +
\frac{1}{m}\,D_{\nu}\left(B_\mu-\frac{1}{2m}D_{\mu}\phi\right).
\eeq{r8} Lagrangian~(\ref{r7}) now enjoys a covariant
St\"{u}ckelberg symmetry:~\footnote{The authors in Ref.~\cite{z2,z3,kliboson}
also considered a gauge invariant description to investigate consistent
interactions of massive high-spin fields. In case of spin-2, say, they
introduce St\"{u}ckelberg fields only in the mass term. This procedure
already breaks St\"{u}ckelberg invariance at tree level. Our approach, instead,
guarantees by construction that St\"{u}ckelberg symmetry is kept intact by the
minimal substitution.} \bea \delta h_{\mu\nu} &=&
D_{\mu}\lambda_{\nu}+D_{\nu}\lambda_{\mu},\label{r9} \\
\delta B_{\mu} &=& D_{\mu}\lambda - m\lambda_{\mu}, \label{r10} \\
\delta\phi &=& 2m\lambda. \eea{r11}

Next we diagonalize the kinetic operators to make sure that the
propagators in the theory have good high energy behavior, i.e., that
all propagators are proportional to $1/p^2$ for momenta $p^2\gg
m^2$. The field redefinition: \beq h_{\mu\nu} \rightarrow h_{\mu\nu}
- \frac{1}{2}\,\eta_{\mu\nu}\phi, \eeq{r14} eliminates the kinetic
mixings between $\phi$ and $h_{\mu\nu}, h$, and also generates a
kinetic term for $\phi$ with the correct sign. After adding the
gauge fixing terms: \bea L_{\text{gf1}}&=& -2\left|D_\nu
h^{\mu\nu}-(1/2)D^\mu h + mB^\mu \right|^2, \label{r16}\\
L_{\text{gf2}}&=&-2\,|D_\mu B^{\mu}+(m/2)(h - 3\phi)|^2,\eea{r17}
we have exhausted all gauge freedom to obtain diagonal kinetic
terms. We are left with \bea L&=&h_{\mu\nu}^{*}(\Box-m^2)h^{\mu\nu}
- \frac{1}{2}h^*(\Box-m^2)h + 2B_\mu^*(\Box-m^2)B^\mu +
\frac{3}{2}\phi^{*}(\Box-m^2)\phi \nonumber \\
&&-\frac{1}{4}F_{\mu\nu}^2 + L_{\text{int}}. \eea{r18} Here $L_{\text{int}}$
contains all interaction operators, which have canonical dimension 4
through 8. Among the non-renormalizable operators the most
potentially dangerous ones, in the high energy limit $m\rightarrow
0$, are the ones with the highest dimensionality. Since in the
decoupling limit $e\rightarrow0$ one is left only with marginal
and relevant operators, any non-renormalizable term must contain
at least one power of $e$. Parametrically in $e\ll1$, for any
given operator dimensionality, the $\mathcal{O}(e)$-terms are more
dangerous than the others. Therefore we are more interested in terms
linear in $e$.

If there exists an $\mathcal{O}(e)$-term that is proportional to the
equations of motion, it can be eliminated by a {\em local} field
redefinition. But this introduces $\mathcal{O}(e^2)$-terms, which
contain even higher dimensional operators. One needs to see if these
$\mathcal{O}(e^2)$-terms be canceled by adding \emph{local}
functions of the high spin field. At $\mathcal{O}(e)$ we have the
following operators of dimensionality 8 and 7: \bea
L_8&=&\frac{e}{m^4}\,\partial_\mu
F^{\mu\nu}[(i/2)\partial_\rho\phi^*
\partial^\rho\partial_\nu\phi+\text{c.c.}]\equiv \frac{e}{m^4}\,\partial_\mu
F^{\mu\nu}J_\nu\label{r19}\\
L_7&=&\frac{ie}{m^3}F^{\mu\nu} \left\{2\partial_\mu
B^*_\rho\partial^\rho\partial_\nu\phi-\partial_\mu B_\nu^* \Box\phi
\right\}+\,\text{c.c.}\,.\eea{r20} The dimension-8 operator can
be removed by the field redefinition: \beq A_\mu\rightarrow
A_\mu-(e/m^4)J_\mu,\eeq{r21} which yields \beq L =
-\frac{1}{4}\,F_{\mu\nu}^2+\frac{e}{m^4}\,\partial_\mu
F^{\mu\nu}J_\nu +~...~,~\rightarrow~-\frac{1}{4}\,F_{\mu\nu}^2 +
\frac{e^2}{4m^8}\, (\partial_\mu J_\nu-\partial_\nu J_\mu)^2
+~...\eeq{r22} This by itself does not improve the degree of
divergence. The field redefinition is helpful, only if we may cancel
the $\mathcal{O}(e^2)$-term in Eq.(\ref{r22}) by adding some
\emph{local} functions of $\tilde{h}_{\mu\nu}$. Such functions may
be present if, for example, there exist other interactions, that are
linear in $F_{\mu\nu}$ and mix the spin-2 field with other more
massive degrees of freedom. By integrating out these additional
degrees of freedom, one ends up with additional EM interactions at
$\mathcal{O}(e^2)$, that involve only the spin-2 field. Indeed the
term \beq L_{\text{add}}=\frac{e^2}{4}(\tilde{h}^*_{\mu\rho}
\tilde{h}^\rho_{~\nu}-\tilde{h}^*_{\nu\rho}\tilde{h}^\rho_{~\mu})^2\eeq{r23}
eliminates the $\mathcal{O}(e^2)$-term in Eq.~(\ref{r22}). In doing
so, we introduce terms proportional to $e^2/m^7$, coming both from
$L_{\text{add}}$, and from the shift of $A_\mu$ acting on the
dimension-7 operator in~(\ref{r20}). We have \beq
L_{11}=\frac{e^2}{2m^7}\,\{\partial^\mu\partial_\sigma\phi^*
\partial^\nu\partial^\sigma\phi-(\mu\leftrightarrow\nu)\}\,\{2\partial_{[\mu}
B^*_{\rho]}\partial^\rho\partial_\nu\phi\ - \partial_{[\mu}
B_{\nu]}^*\Box \phi\}+\text{c.c.}\eeq{r25} We want to eliminate this
dangerous dimension-11 operator up to a total derivative by adding
local counter-terms. In Appendix A using cohomological arguments it
is shown that this is impossible. The key point is that any
\emph{local} function of (ungauged) $\tilde{h}_{\mu\nu}$ is manifestly
invariant under the symmetry: \bea B_\mu&\rightarrow&
B_\mu+b_\mu+b_{\mu\nu}x^\nu, \label{r26}\\\phi&\rightarrow&
\phi+c+c_\mu x^\mu,\eea{r27} where $b_\mu, b_{\mu\nu}$, $c$,
and $c_\mu$ are constants, with $b_{\mu\nu}=-b_{\nu\mu}$. On the
other hand, the dimension-11 operator in~(\ref{r25}) is invariant
under~(\ref{r26}, \ref{r27}) only up to a nontrivial total
derivative\footnote{Under~(\ref{r26}, \ref{r27}) the
\emph{redefined} $U(1)$ field transforms as $\delta
A_\mu=(e/m^4)\delta J_\mu=\partial_\mu
[(ie/2m^4)c_\rho(\partial^\rho\phi -\partial^\rho\phi^*)]$, which is
a total derivative. Since $A_\mu$ appears only in $F_{\mu\nu}$, we
can set $\delta A_\mu=0$.}. As shown in Appendix A, this property is
sufficient to guarantee that $L_{11}$ cannot be canceled up to a
total derivative by local counter-terms. In Appendix B we present a
physical explanation of this fact, by showing that no
local function of $\tilde{h}_{\mu\nu}$ exists, that completely cancels
the contribution of $L_{11}$ to scattering processes.

However, as we will see now, addition of a dipole term leaves us
only with terms proportional to $e/m^3$, which is already an
improvement over field redefinition plus addition of local term.
Indeed, a dipole term $ie\alpha F^{\mu \nu}\tilde{h}^*_{\mu\rho}
\tilde{h}^\rho_{\;\;\nu}$ gives \bea
L_8^{\text{(dipole)}}&=&\frac{e}{m^4}\,\partial_\mu
F^{\mu\nu}[-(i\alpha/2)\partial_\rho\phi^*\partial^\rho\partial_\nu
\phi+\text{c.c.}] \label{r28}\\
L_7^{\text{(dipole)}}&=&-\frac{ie\alpha}{m^3}\,F^{\mu \nu}
\partial_{(\mu} B^*_{\rho)}\partial^\rho\partial_\nu\phi+\,
\text{c.c.}\eea{r29} If we choose $\alpha=1$, in our non-minimal
Lagrangian the dimension-8 operators at $\mathcal{O}(e)$ cancel, and
we are left only with dimension-7 operators: \beq
L_7^{(\text{NM})}=\frac{ie}{2m^3}\,F^{\mu\nu}\left\{2\partial_{[\mu}
B_{\rho]}^*\partial^\rho\partial_\nu\phi-\partial_{[\mu}
B_{\nu]}^*\Box\phi\right\}+\text{c.c.}\eeq{r30} These operators
contain pieces that are not proportional to any of the equations of
motion. Therefore the degree of divergence cannot be improved
further. In the scaling limit: $m\rightarrow0$ and $e\rightarrow0$,
such that $e/m^3$=constant, the non-minimal Lagrangian reduces to:
\beq L=L_{\text{kin}}+\frac{ie}{2m^3}\,F^{\mu\nu}
\left\{2\partial_{[\mu}
B_{\rho]}^*\partial^\rho\partial_\nu\phi-\partial_{[\mu}B_{\nu]}^*
\Box\phi\right\} + \text{c.c.}\eeq{r31} The theory has an intrinsic
cutoff:~\footnote{This is the true cutoff of the theory, while the
``optimistic" one: $\Lambda=m/\sqrt{e}$, mentioned in~\cite{pr1} is
too optimistic.} \beq \Lambda_2= \frac{m}{e^{1/3}}~.\eeq{r32}

It is worth pointing out a couple of interesting facts. First, the
above Lagrangian has acquired a $U(1)$ gauge invariance for the
vector St\"{u}ckelberg field $B_\mu$, i.e., only the field strength of
$B_\mu$ appears in the Lagrangian. This is because an appropriately
chosen dipole term cancels not only the $\mathcal{O}(e)$ dimension-8
operator, but also any $\mathcal{O}(e)$ dimension-7 operators that
are generated by a transformation $\delta
B_{\mu}=\partial_{\mu}\theta$. Second, the dimension-7 operators
in~(\ref{r30}) all cancel for constant $F_{\mu\nu}$. In fact, for
this special case, there exists only one possible dimension-7
operator, and no dimension-8 operators at all. The former, of
course, can be eliminated by appropriately choosing the dipole
coefficient.

\section{Electromagnetically Coupled Massive Spin-3 Field}

The Lagrange formulation for spin-3 fields has been studied
in~\cite{lagspin3}. Their geometric and gauge theoretic aspects
have been discussed in~\cite{gaugespin3}. Electromagnetic and
gravitational interactions of massive spin-3 fields have respectively
been considered in~\cite{emspin3} and~\cite{grspin3}.

For our purpose, what we need is a St\"{u}ckelberg Lagrangian for
massive spin-3 field that can be readily coupled to a $U(1)$ field
or gravity, while maintaining at the same time the covariant version
of the St\"{u}ckelberg symmetry. It can be obtained by the procedure
described below. Let us start with the Lagrangian for a massless
spin-3 field~\cite{ff} in (4+1)D: \beq L = -\frac{1}{2} (\partial_Q
H_{MNP})^2 + \frac{3}{2} (\partial_P H^{MNP})^2
+\frac{3}{2}(\partial_M H_N)^2 + 3H_P\partial_M\partial_N H^{MNP} +
\frac{3}{4}(\partial_M H^M)^2, \eeq{t1} where $H_N=H^M_{~MN}$. The
above Lagrangian has the gauge symmetry: \bea \delta H_{MNP} &=&
\partial_{(M}\Lambda_{NP)},~~~~~~~~\Lambda^M_{~M}=0.\eea{t2}
As we will see, the tracelessness condition on the gauge parameter
has important consequences. Now we do a Kaluza-Klein (KK) reduction
by writing \bea H_{MNP}(x^\mu, x_5) =
\left(\frac{m}{2\pi}\right)^{1/2} \frac{1}{\sqrt{2}}
\left\{h_{MNP}(x^\mu)e^{imx_5} + \text{c.c.}\right\},\eea{t3} where
we compactify the $x_5$-dimension on a circle of radius $1/m$. In
(3+1)D this gives rise to a spin-3 field $h_{\mu\nu\rho}$, a spin-2
field $W_{\mu\nu}\equiv-ih_{\mu\nu5}$, a vector field
$B_\mu\equiv-h_{\mu55}$, and a scalar $\phi\equiv ih_{555}$. We also
write the gauge parameter $\Lambda_{MN}$ as \bea \Lambda_{MN}(x^\mu,
x_5) = \left(\frac{m}{2\pi}\right)^{1/2} \frac{1}{\sqrt{2}}
\left\{\lambda_{MN}(x^\mu)e^{imx_5} + \text{c.c.}\right\},\eea{t4}
so that in (3+1)D we have three gauge parameters:
$\lambda_{\mu\nu}$, $\lambda_\mu\equiv-i\lambda_{\mu5}$, and
$\lambda\equiv-\lambda_{55}$. The higher dimensional gauge
invariance (\ref{t2}) translates itself in lower dimension into the
St\"{u}ckelberg symmetry: \bea \delta h_{\mu\nu\rho} &=&
\partial_{(\mu}\lambda_{\nu\rho)}\label{t5},\\
\delta W_{\mu\nu} &=&
\partial_{(\mu}\lambda_{\nu)}+m\lambda_{\mu\nu},\label{t6}\\
\delta B_{\mu} &=& \partial_{\mu}\lambda + 2m\lambda_{\mu,}\label{t7}\\
\delta\phi &=& 3m\lambda.\eea{t8} The tracelessness of the
5D gauge parameter gives rise to the following condition: \beq
\lambda^\mu_{~\mu}=\lambda. \eeq{t9} We can gauge-fix the KK-reduced
Lagrangian by setting $W^T_{\mu\nu}=0, B_\mu=0$, and $\phi=0$. Note
that because of the constraint (\ref{t9}) only the traceless part
$W^T_{\mu\nu}$ of the spin-2 field $W_{\mu\nu}$ can be set to zero.
This means that the gauge-fixed 4D Lagrangian, which describes a
massive spin-3 field, unavoidably contains an auxiliary scalar field
$W$ $-$ the trace of the would be spin-2 St\"{u}ckelberg field
$W_{\mu\nu}$. We get \bea L &=& -\frac{1}{2}(\partial_\sigma
h_{\mu\nu\rho})^2 + \frac{3}{2} (\partial_\mu h^{\mu\nu\rho})^2 +
\frac{3}{4}(\partial_\mu h^\mu)^2 + \frac{3}{2}(\partial_\mu
h_\nu)^2 + 3h_\rho\partial_\mu\partial_\nu h^{\mu\nu\rho}\nonumber\\
& &-\frac{m^2}{2}(h_{\mu\nu\rho}^2-3h_\mu^2) +
\frac{9}{16}(\partial_\mu W)^2 + \frac{9}{4}m^2W^2 -
\frac{3}{4}m\partial_\mu h^\mu W, \eea{t10} where
$h_\mu=h^\rho_{~\rho\mu}$ is the trace of the spin-3 field. This is
the Lagrangian for a massive spin-3 field with minimal number of
auxiliary fields. After some field redefinitions, it is the same as
the Singh-Hagen spin-3 Lagrangian~\cite{sh,bian}. Considering
the gauge conditions one finds that one can exactly reproduce the
KK-reduced Lagrangian \emph{before} gauge-fixing by the following
field redefinitions:~\footnote{Not surprisingly all the higher
dimensional operators cancel, as required by consistency.} \bea
h_{\mu\nu\rho} \rightarrow ~\tilde{h}_{\mu\nu\rho}&=&h_{\mu\nu\rho} -
\frac{1}{m}\,\partial_{(\mu}W_{\nu\rho)}+ \frac{1}{2m^2}\,
\partial_{(\mu}\partial_{(\nu}B_{\rho))}-\frac{1}{6m^3}\,
\partial_{(\mu}\partial_{(\nu}\partial_{\rho))}\phi\nonumber\\
&&+\frac{1}{4m}\,\eta_{(\mu\nu}\partial_{\rho)}\left(W-\frac{1}{3}
\phi-\frac{1}{m}\,\partial_\sigma B^\sigma + \frac{1}{3m^2}\Box
\phi\right),\label{t11}\\ W \rightarrow ~\tilde{W}&=&W-\frac{1}{3}
\phi -\frac{1}{m}\,\partial_\mu B^\mu + \frac{1}{3m^2}\Box
\phi.\eea{t12} While the Lagrangian~(\ref{t10}) does not have any
manifest St\"{u}ckelberg invariance, after performing the field
redefinitions~(\ref{t11}, \ref{t12}), the St\"{u}ckelberg invariance
is manifest in a trivial manner, because in fact the tensors
$\tilde{h}_{\mu\nu\rho}$ and $\tilde{W}$ themselves are invariant
under the St\"{u}ckelberg transformations~(\ref{t5}-\ref{t8}). The
most important lesson here is that when we couple the theory to a
$U(1)$ field or gravity, the covariant counterparts of the
tensors~(\ref{t11}, \ref{t12}) are still invariant under the
covariant St\"{u}ckelberg transformations. Therefore, for spin-3 we
have been able to construct a consistent Lagrangian that can be
readily coupled to a gauge field, while maintaining at the same time
the covariant version of the St\"{u}ckelberg symmetry.

Now, we couple the massive spin-3 field to electromagnetism by
complexifying the fields in the Lagrangian (\ref{t10}), and
replacing ordinary derivatives with covariant ones
$\partial_\mu\rightarrow D_\mu$: \bea L &=& -|D_\sigma
\tilde{h}_{\mu\nu\rho}|^2 + 3|D_\mu \tilde{h}^{\mu\nu\rho}|^2 +
\frac{3}{2}|D_\mu \tilde{h}^\mu|^2 + 3|D_\mu \tilde{h}_\nu|^2 -
3(D_\mu\tilde{h}_\nu^* D_\rho\tilde{h}^{\mu\nu\rho}+ \text{c.c.})
\nonumber\\
& &-m^2(\tilde{h}^*_{\mu\nu\rho}\tilde{h}^{\mu\nu\rho}
-3\tilde{h}_\mu^*\tilde{h}^\mu) + \frac{9}{8}|D_\mu \tilde{W}|^2 +
\frac{9}{2}m^2\tilde{W}^*\tilde{W} - \frac{3}{4}m(D_\mu
\tilde{h}^\mu \tilde{W}^*+ \text{c.c.}). \eea{t13} Here the
twiddled fields are given by the covariant version of the
tensors~(\ref{t11}, \ref{t12}):
\bea\tilde{h}_{\mu\nu\rho}&=&h_{\mu\nu\rho} -
\frac{1}{m}\,D_{(\mu}W_{\nu\rho)}+
\frac{1}{2m^2}\,D_{(\mu}D_{(\nu}B_{\rho))}
-\frac{1}{6m^3}\,D_{(\mu}D_{(\nu}D_{\rho))}\phi\nonumber\\
&& +\frac{1}{4m}\,\eta_{(\mu\nu}D_{\rho)}\left(W-\frac{1}{3}\phi
-\frac{1}{m}\,D_\sigma B^\sigma + \frac{1}{3m^2}D_\sigma
D^\sigma\phi\right), \label{t14}\\ \tilde{W}&=&W-\frac{1}{3}\phi
-\frac{1}{m}\,D_\mu B^\mu + \frac{1}{3m^2}D_\mu D^\mu\phi.\eea{t15}
One can explicitly work out the various terms in the Lagrangian,
keeping in mind the non-commutativity of covariant derivatives:
$[D_{\mu} , D_{\nu}]=\pm ieF_{\mu \nu}$. The Lagrangian becomes the
sum of three pieces: \beq L = L_{\text{free}} + L_{\text{int}}
-\frac{1}{4}F_{\mu\nu}^2.\eeq{t16} Here $L_{\text{free}}$ is the
free part of the Lagrangian; it consists of kinetic terms, mass
terms, and certain dimension-3 operators, but no higher dimensional
operators. $L_{\text{int}}$ is the interaction Lagrangian, that
consists of various terms, each one containing at least one power of
$e$, and possibly an inverse power of $m\,$($1/m^6$ at most). The
higher dimensional operators appearing here are exactly what we are
interested in. However, we first diagonalize the kinetic terms, by
performing field redefinitions and appropriate gauge fixing, in such
a manner that the propagators in the theory have a good high energy
behavior. We have \bea L_{\text{free}}&=& \{-|\partial_\sigma
h_{\mu\nu\rho}|^2 + 3|\partial_\mu h^{\mu\nu\rho}|^2 +
(3/2)|\partial_\mu h^\mu|^2 + 3|\partial_\mu h_\nu|^2
+3(h_\rho^*\partial_\mu\partial_\nu
h^{\mu\nu\rho}+ \text{c.c.})\nonumber\\
&&-m^2(h^*_{\mu\nu\rho}h^{\mu\nu\rho} -3h_\mu^*h^\mu)
-3[B_\rho^*(\partial_\mu\partial_\nu h^{\mu\nu\rho}-\Box
h^\rho-(1/2)\partial^\rho\partial^\mu h_\mu)+ \text{c.c.}]
\nonumber\\&&+(9/2)|\partial_\mu B^\mu|^2\}+3
\{-|\partial_\sigma W_{\mu\nu}|^2 + 2|\partial_\mu W^{\mu\nu}|^2 +
|\partial_\mu W|^2 +(W^*\partial_\mu\partial_\nu W^{\mu\nu}
+\text{c.c.})\nonumber\\
&&-[\phi^*(\partial_\mu\partial_\nu W^{\mu\nu}-\Box W)+ \text{c.c.}]
+(2/3)|\partial_\mu\phi|^2\}+\frac{3m}{2}\{[2h_{\mu\nu\rho}^*\partial^\mu
W^{\nu\rho} -4h^*_\mu\partial_\nu
W^{\mu\nu}\nonumber\\&&-\partial^\mu h^*_\mu(W-\phi)
-\partial^\mu
B^*_\mu(3W-\phi)]+\text{c.c.}\}+\frac{m^2}{2}|3W-\phi|^2.\eea{t17}
We see that among the dimension-4 kinetic operators, the spin-1 field
$B_\mu$ mixes only with the spin-3 field, $h_{\mu\nu\rho}$, or its
trace, while the spin-0 field $\phi$ mixes only with the spin-2
field, $W_{\mu\nu}$, or its trace. Both kinds of mixing can be
eliminated by standard field redefinitions: \bea h_{\mu\nu\rho}
&\rightarrow& h_{\mu\nu\rho} + \frac{1}{D}\,\eta_{(\mu\nu}B_{\rho)},\\
W_{\mu\nu} &\rightarrow& W_{\mu\nu} + \frac{1}{D-2}\,\eta_{\mu\nu}
\phi.\eea{t18} where $D=4$ is the space-time dimensionality. The
other mixing terms can also be removed by adding to the Lagrangian
the following gauge-fixing terms: \bea
L_{\text{gf1}}&=&-3\,|\,D_\rho h^{\mu\nu\rho}-(1/2)
(D^\mu h^\nu +D^\nu h^\mu) - mW^{\mu\nu} + (m/4)
\eta^{\mu\nu}W\,|^2,\\ L_{\text{gf2}}&=&-6
\,|\,D_\nu W^{\mu\nu} - (1/2)D^\mu W - (m/2)h^\mu -
(5m/4)B^\mu \,|^2,\\
L_{\text{gf3}}&=&-(15/4)\,|\,D_\mu B^\mu - mW -
2m\phi\,|^2.\eea{t19} Note that in adding the term $L_{\text{gf1}}$
we have used up the freedom of a \emph{traceless} symmetric rank-2
gauge parameter. Similarly $L_{\text{gf2}}$ and $L_{\text{gf3}}$
were added at the cost of a vector and a scalar parameter
respectively. This fully fixes all gauge invariances. Now we are
left with a Lagrangian where all the kinetic terms are diagonal:
\bea L&=&h_{\mu\nu\rho}^{*}(\Box-m^2)h^{\mu\nu\rho} -
\frac{3}{2}\,h_{\mu}^{*}(\Box-m^2)h^{\mu}+3W_{\mu\nu}^{*}
(\Box-m^2)W^{\mu\nu}+\frac{15}{4}\,B_{\mu}^{*}(\Box-m^2)
B^{\mu}\nonumber\\&&- \frac{3}{2}\,W^{*}(\Box-m^2)W
+\frac{5}{2}\phi^{*}(\Box-m^2)\phi-\frac{1}{4}\,F_{\mu\nu}^2 +
L_{\text{int}}.\eea{t20} Here all the propagators have the same
pole, which is necessary to cancel spurious poles in tree-level
physical amplitudes.

Now, we turn our attention to the interaction terms, which may
contain higher dimensional operators. Note that with all
the kinetic terms diagonalized, we must assign standard (engineering)
canonical dimensions to the higher-order operators in the
interaction Lagrangian, so that we can interpret ours as an
effective field theory valid up to some cutoff determined by the
most divergent term in the $m\rightarrow 0$ limit. Schematically,
\beq L_{\text{int}}=\sum_{n=4}^{10}L_n.\eeq{t21} Operators with
canonical dimension $n$ are contained in $L_n$; these are multiplied
by a factor $m^{4-n}$. The fact that in the $e\rightarrow0$ limit no
irrelevant operators exist implies that any higher dimensional
operator is at least linear in $e$ (and therefore in $F_{\mu\nu}$).
For fixed $e$, in the high energy limit $m\rightarrow 0$, the higher
the $n$, the more potentially dangerous the operator is. Notice that
the gauge fixing terms can only generate a few harmless, power-counting
renormalizable interactions, that are regular in the massless limit.

The most interesting $\mathcal{O}(e)$ non-renormalizable operators
can be computed in a clever way. To wit: in view
of the powers of $1/m$ appearing in the various terms in the twiddled
fields (\ref{t14}, \ref{t15}), it is clear that the dimension-10 and
-9 operators may arise only from the first line of Lagrangian
(\ref{t13}), which is nothing but the gauged Lagrangian for a
\emph{massless} spin-3 field. We write this as: \beq
L_{\text{massless}}=\tilde{h}_{\mu_1\mu_2\mu_3}^*T^{\mu_1\mu_2\mu_3\nu_1\nu_2\nu_3}
\tilde{h}_{\nu_1\nu_2\nu_3},\eeq{t22} where \bea
T^{\mu_1\mu_2\mu_3\nu_1\nu_2\nu_3}&\equiv&
-3\eta^{\mu_2\nu_2}\eta^{\mu_3\nu_3}D^{\mu_1}D^{\nu_1}
+3\eta^{\nu_1\nu_2}\eta^{\mu_3\nu_3}D^{\mu_1}D^{\mu_2}+3\eta^{\mu_1\mu_2}\eta^{\mu_3\nu_3}D^{\nu_1}
D^{\nu_2}\nonumber\\&&-\frac{3}{2}\,\eta^{\mu_2\mu_3}\eta^{\nu_2\nu_3}D^{\mu_1}D^{\nu_1}
+(\eta^{\mu_1\nu_1}\eta^{\mu_2\nu_2}-3\eta^{\mu_1\mu_2}\eta^{\nu_1\nu_2})\eta^{\mu_3\nu_3}D_\rho
D^\rho,\eea{t23} with symmetrization assumed in
($\mu_1,\mu_2,\mu_3$) and in ($\nu_1,\nu_2,\nu_3$). Keeping in mind
that $D_\mu\equiv\partial_\mu\pm ieA_\mu$, we can expand (\ref{t22})
in powers of $e$: \bea
L_{\text{massless}}&=&\tilde{h}_{\mu_1\mu_2\mu_3}^{*(0)}T^{\mu_1\mu_2\mu_3\nu_1\nu_2\nu_3}_0
\tilde{h}_{\nu_1\nu_2\nu_3}^{(0)}+\tilde{h}_{\mu_1\mu_2\mu_3}^{*(0)}ieT^{\mu_1\mu_2\mu_3\nu_1\nu_2\nu_3}_1
\tilde{h}_{\nu_1\nu_2\nu_3}^{(0)}\nonumber\\&&+~\{ie\tilde{h}_{\mu_1\mu_2\mu_3}^{*(1)}
T^{\mu_1\mu_2\mu_3\nu_1\nu_2\nu_3}_0\tilde{h}_{\nu_1\nu_2\nu_3}^{(0)}+\text{c.c.}\}+\mathcal{O}(e^2),\eea{t24}
where
\bea
T^{\mu_1\mu_2\mu_3\nu_1\nu_2\nu_3}_0&\equiv&
-3\eta^{\mu_2\nu_2}\eta^{\mu_3\nu_3}\partial^{\mu_1}
\partial^{\nu_1}+3\eta^{\nu_1\nu_2}\eta^{\mu_3\nu_3}\partial^{\mu_1}\partial^{\mu_2}+3\eta^{\mu_1\mu_2}
\eta^{\mu_3\nu_3}\partial^{\nu_1}\partial^{\nu_2}\nonumber\\&&-\frac{3}{2}\,\eta^{\mu_2\mu_3}
\eta^{\nu_2\nu_3}\partial^{\mu_1}\partial^{\nu_1}+(\eta^{\mu_1\nu_1}\eta^{\mu_2\nu_2}
-3\eta^{\mu_1\mu_2}\eta^{\nu_1\nu_2})\eta^{\mu_3\nu_3}\Box,\label{t25}\\
\tilde{h}_{\mu\nu\rho}^{(0)}&=&h_{\mu\nu\rho} - \frac{1}{12m}\,\eta_{(\mu\nu}\partial_{\rho)}\phi
-\frac{1}{m}\,\left\{\partial_\mu\left(W_{\nu\rho}-\frac{1}{4}\,\eta_{\nu\rho}W\right)
+\text{cyclic in}~(\mu,\nu,\rho) \right\}\nonumber\\
&&+\frac{1}{m^2}\,\left\{\partial_\mu\left(\frac{1}{2}\,\partial_{(\nu}B_{\rho)}
-\frac{1}{4}\eta_{\nu\rho}\partial_\sigma B^\sigma\right)+\text{cyclic in}~(\mu,\nu,\rho) \right\}
\nonumber\\&&-\frac{1}{3m^3}\,\left\{\partial_\mu\left(\partial_\nu\partial_\rho\phi
-\frac{1}{4}\eta_{\nu\rho}\Box\phi\right)+\text{cyclic in}~(\mu,\nu,\rho)\right\},\label{t26}\\
\nonumber\\ T^{\mu_1\mu_2\mu_3\nu_1\nu_2\nu_3}_1&\equiv& -3\eta^{\mu_2\nu_2}\eta^{\mu_3\nu_3}
(A^{\mu_1}\partial^{\nu_1}-\overleftarrow{\partial}^{\mu_1}A^{\nu_1})
+3\eta^{\nu_1\nu_2}\eta^{\mu_3\nu_3}(A^{\mu_1}\partial^{\mu_2}-\overleftarrow{\partial}^{\mu_1}A^{\mu_2})
\nonumber\\&&+3\eta^{\mu_1\mu_2}\eta^{\mu_3\nu_3}(A^{\nu_1}\partial^{\nu_2}
-\overleftarrow{\partial}^{\nu_1}A^{\nu_2})-\frac{3}{2}\,\eta^{\mu_2\mu_3}\eta^{\nu_2\nu_3}
(A^{\mu_1}\partial^{\nu_1}-\overleftarrow{\partial}^{\mu_1}A^{\nu_1})
\nonumber\\
&&+(\eta^{\mu_1\nu_1}\eta^{\mu_2\nu_2}-3\eta^{\mu_1\mu_2}\eta^{\nu_1\nu_2})\eta^{\mu_3\nu_3}
(A_\rho\partial^\rho-\overleftarrow{\partial}^\rho
A_\rho),
\eea{t27}
and $ie\tilde{h}_{\mu_1\mu_2\mu_3}^{(1)}$
is comprised of the $\mathcal{O}(e)$ terms in
($\tilde{h}_{\mu\nu\rho}-\tilde{h}_{\mu\nu\rho}^{(0)}$). Note that
the quantities inside all the braces in~(\ref{t26}) give vanishing
contribution when $T^{\mu_1\mu_2\mu_3\nu_1\nu_2\nu_3}_0$ (with
proper symmetrization) acts on them, because they are just gauge
shifts of the massless spin-3 field. Thus, in view of
Eq.~(\ref{t26}), only the second term in~(\ref{t24}) can produce
operators of dimensionality 10 and 9 at $\mathcal{O}(e)$. If we are
interested in finding such operators, we need to consider only the
term
$\tilde{h}_{\mu_1\mu_2\mu_3}^{*(0)}ieT^{\mu_1\mu_2\mu_3\nu_1\nu_2\nu_3}_1\tilde{h}_{\nu_1\nu_2\nu_3}^{(0)}$
in~(\ref{t24}). Explicit computation gives: \bea
L_{10}&=&\frac{ie}{m^6}\,F^{\mu\nu}\left[-\partial_\rho\partial_\sigma\partial_\mu\phi^*
\partial^\rho\partial^\sigma\partial_\nu\phi-\frac{7}{24}\,\partial_\mu\Box\phi^*\partial_\nu
\Box\phi
\right].\nonumber\\&=&\frac{e}{m^6}\,\partial_\mu
F^{\mu\nu}\left[\frac{i}{2}\,\partial_\rho\partial_\sigma\phi^*
\partial^\rho\partial^\sigma\partial_\nu\phi+\frac{7i}{48}\,\Box\phi^*\partial_\nu
\Box\phi \right]+\text{c.c.} \eea{t28} \beq
L_9=\frac{ie}{m^5}F^{\mu\nu}\left[3\partial_\mu\partial_\rho
B^*_\sigma
\partial^\rho\partial^\sigma\partial_\nu\phi-\partial_\rho
\partial_\mu B_\nu^*\partial^\rho\Box\phi +\frac{3}{4}\,\Box
B^*_\mu\partial_\nu\Box\phi +\frac{1}{8}\,\partial_\mu
\partial^\sigma B^*_\sigma\partial_\nu\Box\phi\right]+\text{c.c.}\eeq{t29}
The situation is analogous to the spin-2 case, in that the
dimension-10 operator is proportional to the Maxwell equations, so
that it can be removed by a field redefinition of the $U(1)$ gauge
field $A_\mu$. This produces terms proportional to $e^2/m^{12}$,
which can be eliminated by adding of the following local
counter-term to the Lagrangian: \beq L_{\text{add}}=
\frac{e^2}{4}\,\left[(\tilde{h}^*_{\mu\rho\sigma}
\tilde{h}^{\rho\sigma}_{~~\nu}-\tilde{h}^*_{\nu\rho\sigma}
\tilde{h}^{\rho\sigma}_{~~\mu})+\frac{5}{3}(\tilde{h}^*_\mu
\tilde{h}_\nu-\tilde{h}^*_\nu \tilde{h}_\mu)\right]^2.\eeq{t30} In
the process, we end up having $\mathcal{O}(e^2)$ dimension-15
operators: \bea L_{15}&=&\frac{e^2}{2m^{11}}\,\left\{\partial^\mu
\partial_\alpha\partial_\beta \phi^*\partial^\nu\partial^\alpha
\partial^\beta\phi+\frac{7}{24}\,\partial^\mu\Box\phi^*\partial^\nu
\Box\phi-(\mu\leftrightarrow\nu)\right\}\nonumber\\&&~~~~~~~~\times
\{2\partial_\rho\partial_{[\mu}B_{\sigma]}^*\partial^\rho\partial^\sigma
\partial_\nu\phi-\partial_\rho\partial_{[\mu} B_{\nu]}^*\partial^\rho\Box\phi\}
~+~\text{c.c.}\eea{t31} Now any \emph{local} function of the
(ungauged) tensor $\tilde{h}_{\mu\nu\rho}$ enjoys the symmetry: \bea
B_\mu&\rightarrow& B_\mu+b_\mu+b_{\mu\nu}x^\nu+\frac{1}{2}\,
b_{\mu\nu\rho}x^\nu x^\rho,\label{t32}\\
\phi&\rightarrow& \phi+c+c_\mu x^\mu+\frac{1}{2}\,c_{\mu\nu}x^\mu
x^\nu,\eea{t33} where $b_\mu$, $b_{\mu\nu}$,
$b_{\mu\nu\rho}$, $c$, $c_\mu$, and $c_{\mu\nu}$ are constants.
$c_{\mu\nu}$ is symmetric traceless, and $b_{\mu\nu\rho}$ satisfies
$4b_{(\mu\nu\rho)} = b^\alpha_{~\alpha(\mu}\eta^{~}_{\nu\rho)}$.
However, the resulting dimension-15 operators~(\ref{t31}) are
invariant under~(\ref{t32}, \ref{t33}) \emph{only} up to a
nontrivial total derivative. This sets some cohomological
obstruction, shown in Appendix A. In complete analogy with the spin-2
case described in Section 2, they prevent field redefinitions plus
addition of local terms from canceling all terms proportional to
$e^2/m^{11}$.

On the other hand, addition of dipole terms gives us an improved
divergence, where the most divergent terms are proportional to $e/m^5$.
To see this, we notice that at the dipole level we have two possible terms
that respect parity: \beq L_{\text{dipole}} =
ieF^{\mu\nu}\,[\,\beta_1\tilde{h}^*_{\mu\rho\sigma}\tilde{h}_\nu^{~\rho\sigma}
+\beta_2\tilde{h}^*_\mu\tilde{h}_\nu\,].\eeq{t34} They produce the
following $\mathcal{O}(e)$ operators: \bea
L_{10}^{\text{(dipole)}}&=&\frac{ie}{m^6}\,F^{\mu\nu}\left[\beta_1\partial_\rho\partial_\sigma
\partial_\mu\phi^*\partial^\rho\partial^\sigma\partial_\nu\phi+\frac{1}{8}(2\beta_2-\beta_1)\partial_\mu
\Box\phi^*\partial_\nu\Box\phi \right].\label{t35}\\
L_9^{\text{(dipole)}}&=&\frac{ie}{m^5}F^{\mu\nu}
\left[\frac{1}{12}(\beta_1-6\beta_2)\Box B^*_\mu\partial_\nu\Box\phi
+\frac{1}{24}(7\beta_1-6\beta_2)
\partial_\mu\partial^\sigma B^*_\sigma\partial_\nu\Box\phi\right]\nonumber\\&&-\frac{ie\beta_1}{m^5}\,
F^{\mu\nu}\partial_{(\mu}\partial_\rho
B^*_{\sigma)}\partial^\rho\partial^\sigma\partial_\nu\phi
~+~\text{c.c.}\eea{t36} All the dimension-10 operators at
$\mathcal{O}(e)$ cancel in our non-minimal Lagrangian if we set
$\beta_1=1$, and $\beta_2=5/3$. With this choice of dipole
coefficients there are enormous cancelations in dimension-9
operators as well, and we are simply left with: \beq
L_9^{(\text{NM})}=\frac{ie}{2m^5}\,F^{\mu\nu}\left\{2\partial_\rho\partial_{[\mu}B_{\sigma]}^*
\partial^\rho\partial^\sigma\partial_\nu\phi-\partial_\rho\partial_{[\mu} B_{\nu]}^*\partial^\rho
\Box\phi\right\}~+~\text{c.c.} \eeq{t37} Since now the dimension-9
operators contain pieces not proportional to the equations of
motion, we cannot improve the degree of divergence further. In the
scaling limit: $m\rightarrow0$ and $e\rightarrow0$, such that
$e/m^5$=constant, we have the non-minimal Lagrangian: \beq
L=L_{\text{kin}}+\frac{ie}{2m^5}\,F^{\mu\nu}\left\{2\partial_\rho
\partial_{[\mu}B_{\sigma]}^*\partial^\rho\partial^\sigma\partial_\nu
\phi-\partial_\rho\partial_{[\mu}B_{\nu]}^*\partial^\rho\Box
\phi\right\}+\text{c.c.}\eeq{t38} The intrinsic cutoff for our
theory is then given by \beq \Lambda_3=\frac{m}{e^{1/5}}~.\eeq{t39}

We notice that, just as in the case of spin-2, here as well the
Lagrangian has acquired a $U(1)$ gauge invariance for the vector
St\"{u}ckelberg $B_\mu$, i.e., only its field strength shows up in
the Lagrangian~(\ref{t38}). The origin of this gauge invariance can
be understood as follows. When we add the non-minimal (dipole)
terms~(\ref{t34}) to the minimal Lagrangian to cancel all the
$\mathcal{O}(e)$ dimension-10 operators, we effectively change
$T^{\mu_1\mu_2\mu_3\nu_1\nu_2\nu_3}_1$, given in~(\ref{t27}), to
some $T'^{\mu_1\mu_2\mu_3\nu_1\nu_2\nu_3}_1$ such that \bea
0&=&\left\{\partial_{\mu_1}\left(\partial_{\mu_2}\partial_{\mu_3}\phi^*
-\frac{1}{4}\eta_{\mu_2\mu_3}\Box\phi^*\right)+\text{cyclic
in}~(\mu_1,\mu_2,\mu_3) \right\}
ieT'^{\mu_1\mu_2\mu_3\nu_1\nu_2\nu_3}_1\nonumber\\&&\times\left\{\partial_{\nu_1}
\left(\partial_{\nu_2}\partial_{\nu_3}\phi
-\frac{1}{4}\eta_{\nu_2\nu_3}\Box\phi\right)+\text{cyclic
in}~(\nu_1,\nu_2,\nu_3) \right\}\eea{t40} In the above if we
redefine $\phi\rightarrow \phi+\theta$ we must have \bea
0&=&\left\{\partial_{\mu_1}\left(\partial_{\mu_2}\partial_{\mu_3}\theta^*
-\frac{1}{4}\eta_{\mu_2\mu_3}\Box\theta^*\right)+\text{cyclic
in}~(\mu_1,\mu_2,\mu_3) \right\}
ieT'^{\mu_1\mu_2\mu_3\nu_1\nu_2\nu_3}_1\nonumber\\&&\times\left\{\partial_{\nu_1}
\left(\partial_{\nu_2}\partial_{\nu_3}\phi
-\frac{1}{4}\eta_{\nu_2\nu_3}\Box\phi\right)+\text{cyclic
in}~(\nu_1,\nu_2,\nu_3) \right\}~+~\text{c.c.}\eea{t41} Now if the
vector St\"{u}ckelberg $B_\mu$ has a gauge shift $\delta
B_\mu=\partial_\mu\varphi$, it is clear from~(\ref{t26}) that the
gauge shift will result in the following dimension-9 operator: \bea
L_9^{\text{(gauge)}}&\sim&\left\{\partial_{\mu_1}\left(\partial_{\mu_2}
\partial_{\mu_3}\varphi^*-\frac{1}{4}\eta_{\mu_2\mu_3}\Box\varphi^*\right)
+\text{cyclic in}~(\mu_1,\mu_2,\mu_3) \right\} ieT'^{\mu_1\mu_2\mu_3
\nu_1\nu_2\nu_3}_1\nonumber\\&&\times\left\{\partial_{\nu_1}
\left(\partial_{\nu_2}\partial_{\nu_3}\phi-\frac{1}{4}\eta_{\nu_2\nu_3}
\Box\phi\right)+\text{cyclic in}~(\nu_1,\nu_2,\nu_3)
\right\}~+~\text{c.c.}, \eea{t42} which vanishes by virtue
of~(\ref{t41}), if we identify $\theta=-3m\varphi$. Therefore, after
adding the suitable dipole terms to cancel the dimension-10
operators, the dimension-9 operators we are left with must have a
$U(1)$ gauge invariance for the vector St\"{u}ckelberg $B_\mu$.
Similar arguments hold for arbitrary integer spin $s$.

\section{EM Coupling of Massive Rarita-Schwinger Field}

The electromagnetic interaction of massive spin-3/2 is an old
problem. It has been studied by several
authors~\cite{js,vz,spm3half, rs3half,d3,em3half,kli2}. Here we
start with the Lagrangian for a free massive complex
Rarita-Schwinger field: \beq L =
-i\bar{\psi}_\mu\gamma^{\mu\nu\rho}\partial _\nu\psi_\rho -
im\bar{\psi}_\mu\gamma^{\mu\nu}\psi_\nu.\eeq{s1} The mass term does
not enjoy the gauge invariance of the massless part under the
transformation: $\psi_\mu\rightarrow\psi_\mu+\partial_\mu\epsilon$,
where $\epsilon$ is a fermionic gauge parameter. The gauge
invariance is made manifest by introducing a spin-1/2 St\"uckelberg
field $\chi$. When the system is coupled to a $U(1)$ gauge field the
St\"uckelberg invariant Lagrangian reads \beq L =
-i\left(\bar{\psi}_\mu
-\frac{1}{m}\bar{\chi}\overleftarrow{D}_\mu\right)\gamma^{\mu\nu\rho}D_\nu
\left(\psi_\rho-\frac{1}{m}D_\rho\chi\right)-im\left(\bar{\psi}_\mu
-\frac{1}{m}\bar{\chi}\overleftarrow{D}_\mu\right)\gamma^{\mu\nu}\left(\psi_\nu
-\frac{1}{m}D_\nu\chi\right),\eeq{s2} which has the manifest gauged
St\"uckelberg symmetry: \bea
\delta\psi_\mu&=&D_{\mu}\epsilon,\\\delta\chi &=&m\epsilon.\eea{s3}
Working out the Lagrangian (\ref{s2}) one arrives at \bea
L&=&-i\bar{\psi}_\mu\gamma^{\mu\nu\rho}D_\nu\psi_\rho
-im\bar{\psi}_\mu\gamma^{\mu\nu}\psi_\nu + i(\bar{\psi}_\mu
\gamma^{\mu\nu}D_\nu\chi+\bar{\chi}\overleftarrow{D}_\mu\gamma^{\mu\nu}
\psi_\nu)\nonumber\\&&+\,\frac{e}{2m}\,F_{\mu\nu}[\,\bar{\chi}
\gamma^{\mu\nu\rho}\psi_\rho -
\bar{\psi}_\rho\gamma^{\mu\nu\rho}\chi -
\bar{\chi}\gamma^{\mu\nu}\chi - (1/m)\bar{\chi}\gamma^{\mu\nu\rho}
D_\rho\chi\,],\eea{s4} where we have used the non-commutativity of
covariant derivatives: $[D_\mu, D_\nu]=ieF_{\mu\nu}$. The field
redefinition: \beq
\psi_\mu\rightarrow\psi_\mu+\frac{1}{2}\gamma_\mu\chi,\eeq{s5}
eliminates the kinetic mixings, and produces a kinetic term for
$\chi$. The free part of the Lagrangian now becomes \beq
L_{\text{free}}=-i\bar{\psi}_\mu\gamma^{\mu\nu\rho}\partial_\nu\psi_\rho
-\frac{3}{2}\,i\bar{\chi}\not{\!\partial\!}\chi -im\{\bar{\psi}_\mu
\gamma^{\mu\nu}\psi_\nu-3\bar{\chi}\chi+(3/2)(\bar{\psi}_\mu
\gamma^\mu\chi-\bar{\chi}\gamma^\mu\psi_\mu)\}.\eeq{s6} The
$\chi$-propagator is already well-behaved in the $m\rightarrow0$
limit. The spin-3/2 propagator also acquires a good high energy
behavior if we add the gauge fixing term~\cite{n2}: \beq
L_{\text{gf}}=\frac{i}{2}\,\bar{\psi}_\mu\gamma^\mu\gamma^\nu
\gamma^\rho\partial_\nu\psi_\rho. \eeq{s7}

Note that the interaction part of the Lagrangian (\ref{s4}) contains
marginal and irrelevant operators. \emph{After} the shift (\ref{s5})
is made, we find the following terms, singular in the massless limit:
\beq L_{\text{div}}
=\frac{e}{2m}\,F_{\mu\nu}(\bar{\chi}
\gamma^{\mu\nu\rho}\psi_\rho-\bar{\psi}_\rho
\gamma^{\mu\nu\rho}\chi+\bar{\chi}\gamma^{\mu\nu}\chi)-\frac{e}{2m^2}\,F_{\mu\nu}
(\bar{\chi}\gamma^{\mu\nu\rho}\partial_\rho\chi)+\mathcal{O}(e^2).\eeq{s9}
Now we make use of the identity: \beq
\gamma^{\mu\nu\rho}=\gamma^\mu\gamma^\nu
\gamma^\rho-\eta^{\mu\nu}\gamma^\rho+\eta^{\mu\rho}\gamma^\nu-\eta^{\nu\rho}
\gamma^\mu, \eeq{s10} to write the $\mathcal{O}(e)$ dimension-6
operators as \beq L_{6}=\frac{e}{2m^2}\,\partial_\mu
F^{\mu\nu}(\bar{\chi}\gamma_\nu\chi)~-~
\frac{e}{4m^2}\,F_{\mu\nu}(\bar{\chi}\gamma^{\mu\nu}\not{\!\partial\!}\chi-\bar{\chi}
\overleftarrow{\not{\!\partial\!}}\gamma^{\mu\nu}\chi).\eeq{s11} The
entire set is invariant, up to a total derivative, under
$\chi\rightarrow \chi+\xi$, with $\xi =$ constant, which is a
symmetry\footnote{This is true \emph{before} the shift (\ref{s5}) is
made. However, the shift does not yield operators with the highest
dimensionality.} of any local function of
($\psi_\mu-\partial_\mu\chi/m$). We note that addition of any
non-minimal term does not help us improve the degree of divergence,
because any such operator is necessarily irrelevant. For example,
even a dipole term: \beq
L_{\text{dipole}}=\frac{e\alpha}{m}\,F^{\mu\nu}
\left(\bar{\psi}_\mu-\frac{1}{m}\bar{\chi}\overleftarrow{D}_\mu\right)\left(\psi_\nu
-\frac{1}{m}D_\nu\chi\right)\eeq{s12} introduces new dimension-6
operators. Clearly, higher order zero-mass poles can only make it worse. This
does not mean, though, that such operators should not exist. They
do, e.g., in supergravity effective Lagrangians; they simply lower
the EFT cutoff. On the other hand, since all the dimension-6
operators in~(\ref{s11}) are proportional to the equations of
motion, one can cancel them by appropriate field redefinitions of
$A_\mu$ and $\chi$. Indeed \bea A_\mu&\rightarrow& A_\mu
-\frac{e}{2m^2}\,\bar{\chi}\gamma_\mu\chi,\label{s13}\\
\chi&\rightarrow& \chi
+\frac{ie}{6m^2}\,F_{\mu\nu}\gamma^{\mu\nu}\chi,\eea{s14}
serve the purpose, as cancelations occur because of contributions
coming from the kinetic terms. The degree of divergence, at this
point, is not improved though, as now we end up having a bunch of
$\mathcal{O}(e^2)$ dimension-8 operators. We must eliminate all such
operators by adding local functions of
($\psi_\mu-\partial_\mu\chi/m$), if it is possible in the first
place. These operators are of two distinct types. One contains four
$\chi$'s and two derivatives, e.g.,
$\partial_\mu(\bar{\chi}\gamma_\nu\chi)\bar{\chi}\gamma^{\mu\nu}\not{\!\!\partial}\chi$;
the other contains two $\chi$'s, two $F_{\mu\nu}$'s, and one
derivative, e.g., $F_{\mu\nu}
F_{\rho\sigma}\bar{\chi}\gamma^{\mu\nu}\gamma^{\rho\sigma}\not{\!\partial\!}\chi$.
Let us consider the former kind. We need to add a 4-Fermi term of
the spin-3/2 field, like $(e^2/m^2)\bar{\psi}\psi\bar{\psi}\psi$, to
ever get four $\chi$'s. Now each spin-1/2 St\"uckelberg field $\chi$
comes with one derivative and one power of $1/m$, so that we obtain an
$\mathcal{O}(e^2/m^6)$-term with four $\chi$'s, but also with four
derivatives instead of two. Thus the 4-Fermi terms cannot
produce the dangerous operators we wanted to eliminate. We conclude
that the degree of divergence cannot be improved by field
redefinitions and addition of local terms.

Therefore, in the scaling limit: $m\rightarrow0$ and
$e\rightarrow0$, such that $e/m^2$= constant, we have the following
Lagrangian: \beq L=L_{\text{kin}}-\frac{e}{2m^2}\,F_{\mu\nu}
(\bar{\chi}\gamma^{\mu\nu\rho}\partial_\rho\chi).\eeq{s15} It
describes an effective field theory with a finite cutoff \beq
\Lambda_{3/2}=\frac{m}{\sqrt{e}}\,.\eeq{s16} It is the spin-1/2
St\"uckelberg field $\chi$ that becomes strongly coupled at high
energies. The above is the intrinsic cutoff of an
electromagnetically interacting massive Rarita-Schwinger field in
flat space-time background.

\section{Massive Spin-5/2 Coupled to EM}

The field theory of massive and massless spin-5/2 fields has been
discussed in~\cite{n3}. For the massive case, gravitational and EM
interactions were studied in Refs.~\cite{p5half,gr5half} and~\cite{kli2}.

To obtain a St\"{u}ckelberg Lagrangian for massive spin-5/2 field that
can be readily coupled to electromagnetism or gravity, while maintaining
at the same time the covariant version of the St\"{u}ckelberg symmetry,
we start with the Lagrangian for a massless, complex spin-5/2 field~\cite{ff}
in (4+1)D. \bea L &=& -i\,[\,\bar{\Psi}_{MN}\Gamma^A\partial_A\Psi^{MN}
+ 2\bar{\Psi}_{MN}\Gamma^N\Gamma^A\partial_A\Gamma_P\Psi^{PM}
- (1/2)\bar{\Psi}^M_{~M}\Gamma^A\partial_A\Psi^N_{~N}\nonumber\\
&&-2(\bar{\Psi}_{MN}\Gamma^N\partial_A\Psi^{AM}-\text{h.c.})
+ (\bar{\Psi}^M_{~M}\Gamma^A\partial^B\Psi_{AB}-\text{h.c.})\,].\eea{f1}
where $\Gamma^A$'s are 5D Dirac matrices. The Lagrangian enjoys
the gauge symmetry: \beq \delta\Psi_{MN}=\partial_M\Lambda_N
+\partial_N\Lambda_M,~~~~~~~~\Gamma^M\Lambda_M=0.\eeq{f2}
Analogous to the spin-3 case, the condition on the gauge
parameter, namely gamma-tracelessness, has important consequences.
We Kaluza-Klein reduce the
Lagrangian by writing \beq \Psi_{MN}(x^\mu, x_5) = \sqrt{\frac{m}{2\pi}}\,
e^{i(mx_5+\frac{\pi}{4}\gamma_5)}\,\psi_{MN}(x^\mu),\eeq{f3}
where the $x_5$-dimension is compactified on a circle of radius $1/m$.
Here we have incorporated a chiral rotation for convenience. In (3+1)D the
KK-reduction gives rise to a spin-5/2 field $\psi_{\mu\nu}$,
a spin-3/2 field $\xi_\mu\equiv-i\psi_{\mu5}$, and a spin-1/2 field
$\chi\equiv-\psi_{55}$. We also write the gauge parameter $\Lambda_M$ as
\bea \Lambda_M(x^\mu, x_5) = \sqrt{\frac{m}{2\pi}}\,e^{i(mx_5+\frac{\pi}{4}
\gamma_5)}\,\lambda_M(x^\mu).\eea{f4} We have two gauge parameters
in (3+1)D: $\lambda_\mu$, and $\lambda\equiv-i\lambda_5$. The 5D gauge
invariance (\ref{f2}) reduces in lower dimension to the St\"{u}ckelberg
symmetry: \bea \delta\psi_{\mu\nu} &=& \partial_\mu\lambda_\nu
+\partial_\nu\lambda_\mu,\label{f5}\\ \delta\xi_\mu &=&\partial_\mu\lambda
+m\lambda_\mu,\label{f6}\\ \delta\chi &=& 2m\lambda.\eea{f7}
The gamma-tracelessness of the 5D gauge parameter gives the following
condition: \beq \gamma^\mu\lambda_\mu\equiv\not{\!\!\lambda}=\lambda.\eeq{f8}
We gauge fix the KK-reduced Lagrangian by setting $\xi_\mu=(1/4)
\gamma_\mu\not{\!\xi\!}$, and $\chi=0$. Note that the constraint (\ref{f8})
enables us to set to zero only the gamma-traceless part of the spin-3/2 field
$\xi_\mu$. The gauge-fixed 4D Lagrangian, which describes a massive spin-5/2
field, then unavoidably contains an auxiliary spin-1/2 field\,$\not{\!\xi\!}$
$-$ the gamma-trace of the would be spin-3/2 St\"{u}ckelberg field $\xi_\mu$.
One obtains \bea L &=& -i\left[\bar{\psi}_{\mu\nu}\not{\!\partial\!}\psi^{\mu\nu}
+ 2\bar{\psi}_{\mu\nu}\gamma^\nu\not{\!\partial\!}\gamma_\rho\psi^{\rho\mu}
- \frac{1}{2}\bar{\psi}\not{\!\partial\!}\psi-(2\bar{\psi}_{\mu\nu}\gamma^\nu
\partial_\rho\psi^{\rho\mu}-\bar{\psi}\gamma^\mu\partial^\nu\psi_{\mu\nu}-
\text{h.c.})\right]\nonumber\\&&+im\left[\,\bar{\psi}_{\mu\nu}\psi^{\mu\nu}
+ 2\bar{\psi}_{\mu\nu}\gamma^\mu\gamma_\rho\psi^{\rho\nu}-\frac{1}{2}\bar{\psi}\psi\right]
-\frac{i}{4}\left[2\not{\!\bar{\xi}\!}\,\gamma_\mu\partial_\nu\psi^{\mu\nu}+\bar{\psi}
\not{\!\partial\!}\not{\!\xi\!}-\text{h.c.}\right]\nonumber\\
&&-i\not{\!\bar{\xi}\!}\,(\not{\!\partial\!}-m)\not{\!\xi\!}\,,\eea{f9} where
$\psi\equiv\psi^\mu_{~\mu}$ is trace of the spin-5/2 field. This is
the Lagrangian for a massive spin-5/2 field with minimal number of
auxiliary fields. After some field redefinitions it is the same as
the Singh-Hagen spin-5/2 Lagrangian~\cite{sh}. The equivalence of the two
Lagrangians was shown explicitly in~\cite{rs}.

By considering the gauge fixing conditions, we find that we can exactly
reproduce the KK-reduced Lagrangian \emph{before} gauge fixing from the
above Lagrangian, if we make the following field
redefinitions in the latter: \bea \psi_{\mu\nu} ~\rightarrow~
\tilde{\psi}_{\mu\nu}&=&\psi_{\mu\nu} -
\frac{1}{m}\,\partial_{(\mu}\xi_{\nu)}+\frac{1}{2m^2}\,\partial_{(\mu}
\partial_{\nu)}\chi+\frac{1}{4m}\,\gamma_{(\mu}\partial_{\nu)}
\left[\not{\!\xi\!}-\frac{1}{2}\chi-\frac{1}{2m}\not{\!\partial\!}
\chi\right],\label{f10}\\ \not{\!\xi\!}~\rightarrow~
\not{\!\tilde{\xi}\!}&=&\not{\!\xi\!}-\frac{1}{2}\chi-\frac{1}{2m}
\not{\!\partial\!}\chi.\eea{f11} Note that all the higher
dimensional operators cancel, as they should. With these
substitutions, the St\"{u}ckelberg invariance is trivially manifest,
because in fact the tensors $\tilde{\psi}_{\mu\nu}$, and
$\not{\!\tilde{\xi}\!}$\, themselves are invariant under the
St\"{u}ckelberg transformations~(\ref{f5}-\ref{f7}). Most
importantly, when the theory is gauged, the covariant counterparts of
the tensors~(\ref{f10}, \ref{f11}) still preserve the covariant
St\"{u}ckelberg symmetry. Therefore, we have been able to construct
a consistent massive spin-5/2 Lagrangian that can be readily coupled
to a gauge field, while maintaining the covariant version of
St\"{u}ckelberg symmetry.

To couple the theory to electromagnetism we replace ordinary derivatives
with covariant ones: $\partial_\mu\rightarrow D_\mu$, so that
\bea L &=& -i\left[\bar{\tilde{\psi}}_{\mu\nu}\not{\!\!D}\tilde{\psi}^{\mu\nu}
+ 2\bar{\tilde{\psi}}_{\mu\nu}\gamma^\nu\not{\!\!D}\gamma_\rho\tilde{\psi}^{\rho\mu}
- \frac{1}{2}\bar{\tilde{\psi}}\not{\!\!D}\tilde{\psi}-(2\bar{\tilde{\psi}}_{\mu\nu}
\gamma^\nu D_\rho\tilde{\psi}^{\rho\mu}-\bar{\tilde{\psi}}\gamma^\mu D^\nu
\tilde{\psi}_{\mu\nu}-\text{h.c.})\right]\nonumber\\&&
+im\left[\,\bar{\tilde{\psi}}_{\mu\nu}\tilde{\psi}^{\mu\nu}
+ 2\bar{\tilde{\psi}}_{\mu\nu}\gamma^\mu\gamma_\rho\tilde{\psi}^{\rho\nu}
-\frac{1}{2}\bar{\tilde{\psi}}\tilde{\psi}\right]-\frac{i}{4}\left[2\not{\!\bar{\tilde{\xi}}\!}
\,\gamma_\mu D_\nu\tilde{\psi}^{\mu\nu}+\bar{\tilde{\psi}}\not{\!\!D}\not{\!\tilde{\xi}\!}
-\text{h.c.}\right]\nonumber\\&&-i\not{\!\bar{\tilde{\xi}}\!}\,
(\not{\!\!D}-m)\not{\!\tilde{\xi}\!}-\frac{1}{4}F_{\mu\nu}^2, \eea{f12}
where the twiddled fields are the covariant counterparts
of the tensors ~(\ref{f10}, \ref{f11}): \bea \tilde{\psi}_{\mu\nu}&=&
\psi_{\mu\nu} -\frac{1}{m}\,D_{(\mu}\xi_{\nu)}+\frac{1}{2m^2}\,D_{(\mu}
D_{\nu)}\chi+\frac{1}{4m}\,\gamma_{(\mu}D_{\nu)}\left[\not{\!\xi\!}
-\frac{1}{2}\chi-\frac{1}{2m}\not{\!\!D}\chi\right],\label{f13}\\
\not{\!\tilde{\xi}\!}&=&\not{\!\xi\!}-\frac{1}{2}\chi-\frac{1}{2m}
\not{\!\!D}\chi.\eea{f14}
The gauged Lagrangian symbolically looks like: \beq L = L_{\text{free}} + L_{\text{int}}
-\frac{1}{4}F_{\mu\nu}^2.\eeq{f15} $L_{\text{free}}$ is the free part that consists
of kinetic terms, mass terms, and mixed terms, but no higher dimensional operators.
$L_{\text{int}}$ is the interaction Lagrangian, that consists of various terms, each one
containing at least one power of $e$, and having canonical dimensionality 4 through
8. By redefinitions of the fields $\psi_{\mu\nu}$ and $\xi_\mu$, one can get rid of
some of the kinetic mixings. Furthermore we can add suitable gauge fixing terms
(thereby exhausting all gauge freedoms) to the Lagrangian to make sure that the
propagators in the theory have good high energy behavior.
Among others, the above
steps also produce a kinetic term for $\chi$. We do not explicitly carry
out these steps, because they are
not important for the rest of the section. The
important point is that no field redefinition of $\chi$ is needed.

We are interested in finding $\mathcal{O}(e)$-terms containing
operators of the highest possible dimensionality. Note that
dimension-8 operators may only come from spin-5/2 kinetic terms,
i.e., the first line of~(\ref{f12}), which is also the gauged
version of the massless spin-5/2 Lagrangian. We write: \beq
L_{\text{massless}}=-i\bar{\tilde{\psi}}_{\mu\nu}T^{\mu\nu\rho
\sigma\lambda}D_\rho\tilde{\psi}_{\sigma\lambda},\eeq{f16} where
$T^{\mu\nu\rho \sigma\lambda}$ is a constant tensor, symmetric under
$\mu\leftrightarrow\nu$ and $\sigma\leftrightarrow\lambda$, given by
\bea
T^{\mu\nu\rho\sigma\lambda}&=&\frac{1}{2}\,\eta^{\mu\nu}(\eta^{\rho\lambda}
\gamma^\sigma-\eta^{\sigma\lambda}\gamma^\rho)+\frac{1}{2}\,(\eta^{\mu\lambda}
\gamma^{\nu\rho\sigma}+\eta^{\nu\lambda}\gamma^{\mu\rho\sigma})+\frac{1}{2}\,
(\eta^{\sigma\lambda}\eta^{\nu\rho}-\eta^{\rho\lambda}\eta^{\nu\sigma})\gamma^\mu
\nonumber\\&&+\frac{1}{2}\,(\eta^{\sigma\lambda}\eta^{\mu\rho}
-\eta^{\rho\lambda}\eta^{\mu\sigma})\gamma^\nu+\left[\frac{1}{2}
\eta^{\mu\nu}\eta^{\rho\sigma}\gamma^{\lambda}+\frac{1}{2}(\eta^{\mu\sigma}\gamma^{\nu\rho}
+\eta^{\nu\sigma}\gamma^{\mu\rho})\gamma^{\lambda}\right].\eea{f17}
In writing the above we made use of the gamma-matrix
identity~(\ref{s10}). We notice that apart from the terms in the
brackets, $T^{\mu\nu\rho\sigma\lambda}$ is antisymmetric under
$\rho\leftrightarrow\sigma$. We will find this property useful
shortly, as we explore what happens when
$T^{\mu\nu\rho\sigma\lambda}D_\rho$ acts on a gauged St\"{u}ckelberg
shift of $\psi_{\sigma\lambda}$. Armed with the symmetry properties,
it is easy to show that \bea T^{\mu\nu\rho\sigma\lambda}D_\rho
D_{(\sigma}\lambda_{\lambda)}&=&ieF_{\rho\sigma}
\left[\eta^{\mu\nu}\eta^{\rho\lambda}\gamma^{\sigma}+\frac{1}{2}\,(\eta^{\mu\lambda}
\gamma^{\nu\rho\sigma}+\eta^{\nu\lambda}\gamma^{\mu\rho\sigma})+
\eta^{\sigma\lambda}\eta^{\nu\rho}\gamma^\mu+\eta^{\sigma\lambda}\eta^{\mu\rho}
\gamma^\nu\right]\lambda_\lambda\nonumber\\&&+~(\eta^{\mu\nu}\eta^{\rho\sigma}
+\eta^{\mu\sigma}\gamma^{\nu\rho}+\eta^{\nu\sigma}\gamma^{\mu\rho})D_\rho
D_\sigma \not{\!\!\lambda}\eea{f18} where we have used
$[D_\mu,D_\nu]=ieF_{\mu\nu}$. In view of Eq.~(\ref{f13}), we write
\bea \lambda_\mu&=&-\frac{1}{m}\,
\left[\xi_\mu-\frac{1}{4}\gamma_{\mu}\not{\!\xi\!}+\frac{1}{8}\gamma_\mu\chi\right]
+\frac{1}{2m^2}\left[D_\mu\chi-\frac{1}{4}\gamma_{\mu}\not{\!\!D}\chi\right],\label{f19}
\\\text{with}~~~\not{\!\!\lambda}&=&-\frac{1}{2m}\,\chi. \eea{f20}

Now that all possible dimension-8 operators coming from the
Lagrangian~(\ref{f12}) are contained only in
$\bar{\lambda}_{(\mu}\overleftarrow{D}_{\nu)}T^{\mu\nu\rho\sigma\lambda}
D_\rho D_{(\sigma}\lambda_{\lambda)}$, where $\lambda_\mu$ is given
by~(\ref{f19}), one can easily find them all. Note that the terms on
the second line of Eq.~(\ref{f18}) do not contribute the dimension-8
operators, because they can at most produce dimension-7 operators.
Therefore at $\mathcal{O}(e)$ we find the following operators: \bea
L_8&=&-\frac{i}{4m^4}\,(2ieF_{\rho\sigma})\left\{\bar{\chi}\overleftarrow{\partial}_\nu
\overleftarrow{\partial}_\mu-\frac{1}{4}\,\bar{\chi}\overleftarrow{\not{\!\partial\!}}\gamma_\nu
\overleftarrow{\partial}_\mu\right\}\nonumber\\&&\times
\left[\eta^{\mu\nu}
\eta^{\rho\lambda}\gamma^{\sigma}+\frac{1}{2}\,(\eta^{\mu\lambda}\gamma^{\nu\rho\sigma}
+\eta^{\nu\lambda}\gamma^{\mu\rho\sigma})+\eta^{\sigma\lambda}(\eta^{\nu\rho}\gamma^\mu
+\eta^{\mu\rho}\gamma^\nu)\right]\left\{{\partial}_\lambda\chi-\frac{1}{4}\,\gamma_\lambda
\not{\!\partial\!}\chi\right\}.\nonumber\eea{xxx1} Notice that the
double-derivative tensor appearing in the braces above is zero under
contraction with $\gamma^\nu$. One can simplify the above by making
use of various identities and (anti)commutation relations involving
gamma matrices and products thereof, Bianchi identity etc. to
obtain: \beq L_8=-\frac{e}{2m^4}\,F_{\mu\nu}
(\bar{\chi}\overleftarrow{\partial_\sigma}\gamma^{\mu\nu\rho}\partial_\rho\partial^\sigma\chi)
+\frac{3e}{16m^4}\,F_{\mu\nu}(\bar{\chi}\overleftarrow{\not{\!\partial\!}}\gamma^{\mu\nu\rho}
\partial_\rho\not{\!\partial\!}\chi).\eeq{f21} Each term here is invariant
up to a total
derivative under $\chi\rightarrow \chi+\varepsilon+ \varepsilon_\mu
x^\mu$, where $\varepsilon$ is a constant spinor, and
$\varepsilon_\mu$ a constant vector-spinor. In fact, the above local
transformation is a symmetry of the highest dimensional operators
built out of any local function of the high spin field. Addition of
any non-minimal terms at this point does not improve the degree of
divergence; they rather make it worse by introducing new dimension-8
operators (see remarks in section 4). However, by using the
identity~(\ref{s10}), one can render the operators in
Eq.~(\ref{f21}) proportional to the equations of motion: \bea
L_8&=&\frac{e} {2m^4}\partial_\mu
F^{\mu\nu}\left(\bar{\chi}\overleftarrow{\partial_\sigma}\gamma_\nu\partial^\sigma
\chi-\frac{3}{8}\bar{\chi}\overleftarrow{\not{\!\partial\!}}\gamma_\nu\not{\!\partial\!}\chi\right)
-\frac{e}{4m^4}\left(\bar{\chi}\overleftarrow{\partial_\sigma}\not{\!\!F}\partial^\sigma
\not{\!\partial\!}\chi-\frac{3}{8}\bar{\chi}\overleftarrow{\not{\!\partial\!}}\not{\!\!F}\Box\chi
+...\right)\nonumber\\&&~~\eea{f22} where $\not{\!\!\!F} \equiv
F_{\mu\nu}\gamma^{\mu\nu}$, and (...) stands for hermitian
conjugate. We are in a situation analogous to the spin-3/2 case.
Although by appropriate field redefinitions of $A_\mu$ and $\chi$
one may cancel all these terms, the $\mathcal{O}(e^2)$ dimension-12
operators obtained in the process cannot be eliminated by adding
local functions of the high spin field (one can see it just by
considering terms containing four $\chi$'s). In other words, the
degree of divergence cannot be improved by field redefinitions plus
addition of local terms.

In the scaling limit: $m\rightarrow0$ and $e\rightarrow0$, such that $e/m^4$=constant, our
Lagrangian reduces to: \beq L=L_{\text{kin}}-\frac{e}{2m^4}\,F_{\mu\nu}(\bar{\chi}\overleftarrow{
\partial_\sigma}\gamma^{\mu\nu\rho}\partial_\rho\partial^\sigma\chi)+\frac{3e}{16m^4}\,F_{\mu\nu}
(\bar{\chi}\overleftarrow{\not{\!\partial\!}}\gamma^{\mu\nu\rho}\partial_\rho\not{\!\partial\!}
\chi).\eeq{f23} Thus we have an effective field theory with an
intrinsic finite cutoff \beq \Lambda_{5/2}=\frac{m}{e^{1/4}}\,.\eeq{f24}
It is again the spin-1/2 St\"uckelberg field $\chi$ that plays the principal
role, by being the strongest interacting mode at high energies.

\section{Intrinsic Cutoff for Arbitrary Spin}

Having explicitly worked out the examples of spin 2, 3, 3/2, and
5/2, we see a generic pattern in the expression for the intrinsic
cutoff of the theory as a function of the particle's mass, spin, and
electric charge. One is tempted to conjecture that for any spin-$s$
particle of mass $m$ and electric charge $e$, the parametric
dependence of the cutoff on $m$ and $e$ is
\beq
\Lambda_s={\cal O}\left(\frac{m}{e^{1/(2s-1)}}\right)\,.
\eeq{g1}
We will now show
that the above is indeed the expression for the upper bound on the cutoff.
To do
so we consider the cases of integer spin and half-integer spin
separately.

\subsection{Integer Spin $s$:}

One can as usual follow the procedure described in the introduction.
We outline steps 1 through 4 only briefly, because the details are
not very important for our final conclusion. One starts with the
massless Lagrangian in (4+1)D~\cite{ff}, and then Kaluza-Klein
reduce it to obtain in (3+1)D a St\"uckelberg invariant Lagrangian
for a massive spin-$s$ field. The KK-reduced Lagrangian before gauge
fixing contains symmetric tensor fields of rank (spin) $0$ through
$s$; the St\"uckelberg
symmetry reads: \bea \delta\phi_s&=&\partial\lambda_{s-1}\nonumber\\
\delta\phi_{s-1}&=&\partial\lambda_{s-2}+m\lambda_{s-1}\nonumber\\
&...&\nonumber\\\delta\phi_{s-k}&=&\partial\lambda_{s-k-1}
+km\lambda_{s-k}\nonumber\\&...&\nonumber\\\delta\phi_{0}&=&
sm\lambda_{0}\nonumber\eea{xxx2} The tracelessness of the 5D gauge
parameter gives rise to ($s-2$) conditions on the 4D gauge
parameters: \bea \lambda_{s-1}'&=&\lambda_{s-3}\nonumber\\
\lambda_{s-2}'&=&\lambda_{s-4}\nonumber\\&...&\nonumber\\\lambda_2'&=&
\lambda_0\nonumber\eea{xxx3} where prime denotes trace w.r.t. the Minkowski
metric. Thus, we have at our disposal $s$ symmetric, \emph{traceless},
independent
gauge parameters: $\{\lambda_0, \lambda_1, \lambda_{2}^T, ...,
\lambda_{s-1}^T\}$. This implies that the
gauge fixed 4D Lagrangian not only contains a spin-$s$ field
$\phi_s$ and its trace $\phi_s'$ (spin $s-2$), but also ($s-2$)
auxiliary fields with spins 0 through ($s-3$): $\{\phi_2', \phi_3',
..., \phi_{s-1}'\}$, identified as traces of St\"uckelberg fields.
One can construct the following symmetric tensors that are invariant
under the St\"uckelberg transformations. \bea \hat{\phi}_s&=&
\sum_{k=0}^s\frac{(-1)^k}{k!\,m^k}\,\partial^k\phi_{s-k},\label{b1}\\
\hat{\phi}'_n&=&\phi'_n-(n-2)\phi_{n-2}-\left(\frac{2/m}{s-n+1}\right)
\partial.\phi_{n-1}\nonumber\\&&+\sum_{k=2}^n
\frac{(-1)^k(s-n)!}{(s-n+k)!\,m^k}\,[\,m^2a(s,n)+2\partial\cdot\partial\,]\,
\partial^{k-2}\phi_{n-k},\eea{b2} \\ where $n = 2, 3, ..., (s-1)$,
and $a(s,n)\equiv\{s(n-3)-(n-1)(n-4)\}(s-n+1)$. The construction of
the ($s-2$) lower rank invariant tensors~(\ref{b2}) has been
possible because of the ($s-2$) gauge conditions. Simplest among
them is the invariant scalar: \beq
\hat{\phi}'_2=\phi'_2-\left(\frac{s-2}{s}\right)
\phi_0-\frac{2/m}{s-1}\,\left[\partial\cdot\phi_1-\frac{1}{sm}\,\Box
\phi_0\right]. \eeq{b2zero} Consider the following field
redefinitions in the gauge fixed Lagrangian:
\bea \phi_2'~\rightarrow\tilde{\phi}_2'&=& \hat{\phi}_2'\label{b3}\\
\phi_3'~\rightarrow
\tilde{\phi}_3'&=&\hat{\phi}_3'+\left(\frac{2}{s-2}\right)\,\frac{1}{4m}\,
\partial\hat{\phi}_2' \label{b4}\\ &...& \nonumber\\
\phi_s~\rightarrow\tilde{\phi}_s&=&\hat{\phi}_s+\frac{1/m}{2(s-1)}\,
\eta\,\partial\left[\hat{\phi}_{s-1}'+\sum_{n=2}^{s-2}\frac{b(s)}{m^{n-1}}\,
\partial^{n-1}\hat{\phi}'_{s-n}\right],\eea{b5} where $b(s)$ is
some rational function of $s$. Once~(\ref{b3}-\ref{b5}) are
performed, the St\"uckelberg invariant KK-reduced Lagrangian
\emph{before} gauge fixing is reproduced. One can understand this
fact by considering the gauge fixing conditions. The function $b(s)$
can be determined as follows. Note that $\tilde{\phi}_s$ and
$\phi_s$ are related by the gauge transformation: \beq
\tilde{\phi}_s=\phi_s+\partial\lambda_{s-1}
=\phi_s+\partial\lambda_{s-1}^T+\frac{1}{4}\,\eta\,\partial\lambda_{s-3}.\eeq{b6}
We can compare the coefficients of the terms containing $(1/m)^s$
in~(\ref{b5}) and~(\ref{b6}). Our gauge fixing conditions are such
that $\lambda_{k-1}$ contains at most $(1/m)^k$. Thus we only need
to consider the term $\partial\lambda_{s-1}^T$ in~(\ref{b6}). In
view of~(\ref{b1}), the tracelessness of $\lambda_{s-1}^T$ gives
\beq \tilde{\phi}_s=\phi_s+\frac{(-1)^s}{s!\,m^s}
\left[\partial^s\phi_0-\frac{2(s-1)}{(s-1)(s-2)+6}\,\eta\,
\partial^{s-2}\Box\phi_0 \right]+~...\eeq{b7} By comparing the
above with Eq.~(\ref{b5}), we can find $b(s)$.

The virtue of the above procedure is that the St\"uckelberg
invariance is left intact by the minimal substitution
$\partial_\mu\rightarrow D_\mu$. The gauged Lagrangian with explicit
Lorentz indices is given by \bea L&=&-|D_\rho\tilde{\phi}_{\mu_1\mu_2
...\mu_s}|^2 +s|D^{\mu_1}\tilde{\phi}_{\mu_1\mu_2...\mu_s}|^2
+\frac{s(s-1)}{2}\,|D_\rho\tilde{\phi}^{\mu_1}_{~\mu_1\mu_3...
\mu_s}|^2\nonumber\\&&+\frac{s(s-1)}{2}\,[D^{\mu_1}D^{\mu_2}
\tilde{\phi}^*_{\mu_1\mu_2...\mu_s}\tilde{\phi}_{\nu_1}^{~\nu_1\mu_3...\mu_s}
+\text{c.c.}]+\frac{s(s-1)(s-2)}{4}\,|D^{\mu_3}\tilde{\phi}^{\mu_1}_{~\mu_1\mu_3
...\mu_s}|^2\nonumber\\&&-m^2\left(|\tilde{\phi}_{\mu_1\mu_2...\mu_s}|^2
-\frac{s(s-1)}{2}\,|\tilde{\phi}^{\mu_1}_{~\mu_1\mu_3...\mu_s}|^2\right)+~(...)~
-\frac{1}{4}F_{\mu\nu}^2,\eea{b8} where the twiddled fields are given
by the covariant counterparts of~(\ref{b3}-\ref{b5}), and the
ellipses (...) stand for terms involving lower spin (auxiliary)
fields. Non-commutativity of covariant derivative will give rise to
higher dimensional interaction operators, that contain powers of
$e$, in the above Lagrangian. Symbolically: \beq L = L_{\text{free}}
+ L_{\text{int}} -\frac{1}{4}F_{\mu\nu}^2.\eeq{b9} By doing
appropriate field redefinitions and adding gauge fixing terms
(thereby exhausting all gauge freedom) we can produce diagonal
kinetic operators for all the fields, so that the propagators in the
theory have good high energy behavior. It is important to note that
the spin-1 and spin-0 St\"uckelberg fields do not require any field
redefinition.

The interaction Lagrangian may contain operators up to
dimension ($2s+4$). Operators of mass dimension ($2s+4$) and
($2s+3$) may only come from the first five terms of the
Lagrangian~(\ref{b8}); these terms comprise the gauged Lagrangian
for a \emph{massless} spin-$s$ field. An analysis similar to that
presented for spin-3 shows that \beq L_{\text{int}}=
\tilde{\phi}_{\mu_1\mu_2...\mu_s}^{*(0)}ieT^{\mu_1\mu_2...\mu_s;\nu_1\nu_2...
\nu_s}_1\tilde{\phi}_{\nu_1\nu_2...\nu_s}^{(0)}+L_{\leq(2s+2)}
+\mathcal{O}(e^2),\eeq{b10} where $\tilde{\phi}_{\mu_1\mu_2...
\mu_s}^{(0)}$ is given by~(\ref{b5}) or~(\ref{b7}), and \bea
T^{\mu_1\mu_2...\mu_s;\nu_1\nu_2...\nu_s}_1&\equiv& \left\{\eta^{
\mu_1\nu_1} \eta^{\mu_2\nu_2}-\frac{s(s-1)}{2}\,\eta^{\mu_1\mu_2}
\eta^{\nu_1\nu_2}\right\}\eta^{\mu_3\nu_3}...\eta^{\mu_s \nu_s}
(A_\rho\partial^\rho-\overleftarrow{\partial}^\rho A_\rho)\nonumber\\
&&+~\frac{s(s-1)}{2}\,\eta^{\mu_3\nu_3}...\eta^{\mu_s\nu_s}[\eta^{\mu_1
\mu_2}(A^{\nu_1}\partial^{\nu_2}-\overleftarrow{\partial}^{\nu_1}A^{\nu_2})
+(\mu_i\leftrightarrow\nu_i)]\nonumber\\&&-~\frac{s(s-1)(s-2)}{4}\,
\eta^{\mu_4\nu_4}...\eta^{\mu_s\nu_s}\eta^{\mu_2\mu_3}\eta^{\nu_2\nu_3}
(A^{\mu_1}\partial^{\nu_1}-\overleftarrow{\partial}^{\mu_1}
A^{\nu_1})\nonumber\\&&-~s\eta^{\mu_2\nu_2}...\eta^{\mu_s\nu_s}
(A^{\mu_1}\partial^{\nu_1}-\overleftarrow{\partial}^{\mu_1}
A^{\nu_1}), \eea{b11} with symmetrization assumed in ($\mu_1,\mu_2,
..., \mu_s$) and in ($\nu_1,\nu_2, ..., \nu_s$). Given Eq.~(\ref{b10}),
it is not difficult to compute the dimension-($2s+4$) and -($2s+3$)
operators at $\mathcal{O}(e)$. Let us denote the spin-0
St\"uckelberg field as $\phi_0\equiv\phi$, and the spin-1 one as
$\phi_1\equiv B_\mu$. We have \bea L_{2s+4}&=&-~\frac{ie}{m^{2s}}\,
F^{\mu\nu}\partial_{\mu_1}...\partial_{\mu_{s-1}}\partial_\mu
\phi^*\partial^{\mu_1}...\partial^{\mu_{s-1}}\partial_\nu
\phi\nonumber\\ &&+~\frac{ie}{m^{2s}}\,c_1(s)F^{\mu\nu}
\partial_{\mu_1}...\partial_{\mu_{s-3}}\partial_\mu\Box\phi^*
\partial^{\mu_1}...\partial^{\mu_{s-3}}\partial_\nu\Box
\phi,\label{b12}\\ L_{2s+3}&=&\frac{ie}{m^{2s-1}}\,s
F^{\mu\nu}\partial_\mu\partial_{\mu_1}...\partial_{\mu_{s-2}}B_\rho^*
\partial^{\mu_1}...\partial^{\mu_{s-2}}\partial^\rho\partial_\nu \phi+
\text{c.c.}\nonumber\\&&-~\frac{ie}{m^{2s-1}}\,F^{\mu\nu}
\partial_{\mu_1}...\partial_{\mu_{s-2}}\partial_\mu B_\nu^*\partial^{\mu_1}
...\partial^{\mu_{s-2}}\Box\phi+\text{c.c.}\nonumber\\&&+~
\frac{ie}{m^{2s-1}}\,c_2(s)F^{\mu\nu}\partial_{\mu_1}...\partial_{\mu_{s-3}}
\Box B_\mu^*\partial^{\mu_1}...\partial^{\mu_{s-3}}\partial_\nu\Box
\phi+\text{c.c.} \nonumber\\&&+~\frac{ie}{m^{2s-1}}\,
c_3(s)F^{\mu\nu}\partial_{\mu_1} ...\partial_{\mu_{s-3}}\partial_\mu
(\partial^\sigma B_\sigma^*)\partial^{\mu_1}...\partial^{\mu_{s-3}}
\partial_\nu\Box\phi+\text{c.c.},\eea{b13}
where the $c(s)$'s are rational functions of $s$, which are not
important in the subsequent discussion. Here the dimension-($2s+4$)
operators are proportional to the Maxwell equations: $L_{2s+4}=
(e/m^{2s})\partial_\mu F^{\mu\nu}J_\nu$, where $J_\nu$ is given by
\beq J_\nu\equiv\frac{i}{2}\,\left[\partial_{\mu_1}
...\partial_{\mu_{s-1}}\phi^*\partial^{\mu_1}...\partial^{\mu_{s-1}}
\partial_\nu\phi-\,c_1(s)\partial_{\mu_1}...\partial_{\mu_{s-3}}
\Box\phi^*\partial^{\mu_1}...\partial^{\mu_{s-3}}\partial_\nu\Box
\phi\right]+\text{c.c.} \eeq{b12prop} Therefore, they can be eliminated
by a field redefinition of the $U(1)$ gauge field. The resulting
$\mathcal{O}(e^2)$ dimension-($4s+4$) operators can be canceled by
addition of local counter-terms of the spin-$s$ field. This leaves
us with terms proportional to $e^2/m^{4s-1}$.

We expect here a situation similar to that encountered
in the cases of spin-2 and spin-3.
Any \emph{local} function of the (ungauged) tensor $\tilde{\phi}_s$
is fully invariant under the symmetry: \bea B_\mu
&\rightarrow& B_\mu
+b_\mu+\sum_{k=1}^{s-1}\frac{1}{k!}\,b_{\mu,\nu_1...\nu_k}x^{\nu_1}...
x^{\nu_k},\label{b14}\\\phi&\rightarrow&
\phi+c+\sum_{k=1}^{s-1}\frac{1}{k!}\,c_{\mu_1...\mu_k}x^{\mu_1}...
x^{\mu_k},\eea{b15} where $b_\mu$, $b_{\mu,\nu_1...\nu_k}$,
$c$, and $c_{\mu_1...\mu_k}$ are constants. $c_{\mu_1...\mu_k}$ is
symmetric traceless; $b_{\mu,\nu_1...\nu_k}$, which is symmetric
under ($\nu_i\leftrightarrow\nu_j$) by definition, satisfies for
$k=(s-1)$ the following: $b_{(\mu,\nu_1...\nu_{s-1})}= 0$,
$b^\mu_{~\mu\nu_2...\nu_{s-1}}=0$~\footnote{This set of conditions
is actually stronger than necessary. But as long as
$b_{\mu,\nu_1...\nu_{s-1}}\neq 0$, it does not matter for our
purpose.}. Given that the dimension-($4s+3$) operators obey
this symmetry only up to a nontrivial total derivative,
the cohomological argument presented in Appendix A should prevent
us from improving the degree of divergence any further. No new feature
needed to the computation of the cohomology arises above spin-3, except for a
clutter of Lorentz indices. So, we will assume that the cohomological
obstruction present in the case of spin-2 and 3 exists for all integer spins.

However, one can always add to the minimal Lagrangian~(\ref{b8}) a
dipole term, which depends on two free parameters $\beta_1$ and
$\beta_2$\footnote{Terms involving
$\tilde{F}$ are not useful in canceling the existing
non-renormalizable operators, because the two have opposite parity.}:
\beq L_{\text{dipole}}=
ieF^{\mu\nu}\, [\,\beta_1\tilde{\phi}^*_{\mu\mu_2...\mu_s}
\tilde{\phi}_\nu^{~~\mu_2 ...\mu_s}+\beta_2\,\tilde{\phi}_{\mu
\mu_2}^{*~~~\mu_2\mu_4...\mu_s} \tilde{\phi}_{\nu~~\nu_2\mu_4...
\mu_s}^{~~\nu_2}\,].\eeq{b16}
We get \bea
L_{2s+4}^{\text{(dipole)}}&=&\frac{ie}{m^{2s}}\,\beta_1
F^{\mu\nu}\partial_{\mu_1}...\partial_{\mu_{s-1}}\partial_\mu
\phi^*\partial^{\mu_1}...\partial^{\mu_{s-1}}\partial_\nu
\phi\nonumber\\&&+~\frac{ie}{m^{2s}}\,f_1(\beta_1,\beta_2)
F^{\mu\nu}\partial_{\mu_1}...\partial_{\mu_{s-3}}\partial_\mu\Box
\phi^*\partial^{\mu_1}...\partial^{\mu_{s-3}}\partial_\nu\Box
\phi,\label{b17}\\L_{2s+3}^{\text{(dipole)}}
&=&-~\frac{ie}{m^{2s-1}}\, \beta_1 F^{\mu\nu}\partial_{(\mu}
\partial_{\mu_1}...\partial_{\mu_{s-2}}B_{\rho)}^*\partial^{\mu_1}...
\partial^{\mu_{s-2}}\partial^\rho\partial_\nu\phi+
\text{c.c.}\nonumber\\&&+~\frac{ie}{m^{2s-1}}\,
f_2(\beta_1,\beta_2)F^{\mu\nu}\partial_{\mu_1}...\partial_{\mu_{s-3}}
\Box B_\mu^*\partial^{\mu_1}...\partial^{\mu_{s-3}}\partial_\nu
\Box\phi +\text{c.c.}\nonumber\\&&+~\frac{ie}{m^{2s-1}}\,
f_3(\beta_1,\beta_2)F^{\mu\nu}\partial_{\mu_1}...\partial_{\mu_{s-3}}
\partial_\mu(\partial^\sigma B_\sigma^*)\partial^{\mu_1}...
\partial^{\mu_{s-3}}\partial_\nu\Box\phi+\text{c.c.},\eea{b18}
where $f(\beta_1,\beta_2)$'s are linear functions of $\beta_1,
\beta_2$. The $\mathcal{O}(e)$ dimension-($2s+4$) operators can be
canceled by choosing $\beta_1=1$, and $\beta_2$ such that
$f_1(1,\beta_2)+c_1(s)=0$. As discussed in Section 3, this
will give rise to a gauge invariance for the vector St\"{u}ckelberg
in the $\mathcal{O}(e)$ dimension-($2s+3$) operators. Since all the
operators that do not respect this gauge invariance cancel, we must
have $f_2(1,\beta_2) +c_2(s)=0$, and $f_3(1,\beta_2)+c_3(s)=0$.

Now in the scaling limit: $m\rightarrow0$ and $e\rightarrow0$, such
that $e/m^{2s-1}$=constant, the non-minimal Lagrangian reduces to
\bea L&=&L_{\text{kin}}+\frac{ieF^{\mu\nu}}{m^{2s-1}}\,
\{\partial_{\mu_1}...\partial_{\mu_{s-2}}\partial_{[\mu}
B_{\rho]}^*\partial^{\mu_1}...\partial^{\mu_{s-2}}\partial^\rho
\partial_\nu\phi-\text{c.c.}\}\nonumber\\&&-~
\frac{ieF^{\mu\nu}}{2m^{2s-1}}\,\{\partial_{\mu_1}...
\partial_{\mu_{s-2}}\partial_{[\mu} B_{\nu]}^*\partial^{\mu_1}...
\partial^{\mu_{s-2}}\Box\phi-\text{c.c.}\}\eea{b19} Since some of
the dimension-($2s+3$) operators are not proportional to any of the
equations of motion, we cannot improve the degree of
divergence. Thus the theory has an intrinsic UV cutoff, not higher
than \beq \Lambda_s=\frac{m}{e^{1/(2s-1)}}~.\eeq{b20}

\subsection{Half-integer Spin $s=n+1/2$:}

We start with the Lagrangian for a massless field of arbitrary
half-integer spin $s=n+1/2$ in (4+1)D~\cite{ff}. Kaluza-Klein
reduction to (3+1)D gives a St\"uckelberg invariant Lagrangian for a
massive spin-$s$ field. The Lagrangian contains symmetric
tensor-spinor fields of rank $0$ through $n$ (spin $1/2$ through
$s$), and enjoys the St\"uckelberg symmetry: \bea \delta\psi_n&=&
\partial\lambda_{n-1}\nonumber\\ \delta\psi_{n-1}&=&\partial\lambda_{n-2}
+m\lambda_{n-1}\nonumber\\&...&\nonumber\\\delta\psi_{n-k}&=&\partial
\lambda_{n-k-1}+km\lambda_{n-k}\nonumber\\&...&\nonumber\\ \delta
\psi_{0}&=&nm\lambda_{0}\nonumber\eea{xxx4} The 4D gauge parameters are
not all independent; they satisfy the following ($n-1$) conditions,
thanks to the gamma-tracelessness of the 5D gauge parameter:
\bea \lambda_{n-1}'&=&\lambda_{n-2}\nonumber\\
\lambda_{n-2}'&=&\lambda_{n-3}\nonumber\\&...&\nonumber\\
\lambda_1'&=& \lambda_0\nonumber\eea{xxx5} where prime denotes
gamma-trace. We therefore have the freedom of $n$ symmetric
\emph{gamma-traceless} independent gauge parameters, namely
$\{\lambda_0, \lambda_1^{\gamma T}, \lambda_{2}^{\gamma T}, ...,
\lambda_{n-1}^{\gamma T}\}$. The gauge fixed Lagrangian contains the
following fields $-$ a spin-$s$: $\psi_n$, a spin-$(s-1)$: $\psi'_n$,
two spin-$(s-2)$: $\psi''_n, \psi'_{n-1}$, two spin-$(s-3)$:
$\psi''_{n-1}, \psi'_{n-2}$, ... ..., two spin-1/2: $\psi''_2,
\psi'_1$. All the auxiliary fields have been identified as
(gamma)traces of the St\"uckelberg fields.

The following tensor-spinor is invariant under the St\"uckelberg
transformation: \beq \hat{\psi}_n\equiv\sum_{k=0}^n
\frac{(-1)^k}{k!\,m^k}\,\partial^k\psi_{n-k}. \eeq{d1} Because of
the $(n-1)$ conditions on the gauge parameters, one is also able to
construct $(n-1)$ additional invariant tensor-spinors: $\hat{\psi}'_1,
\hat{\psi}'_2, ..., \hat{\psi}'_{n-1}$. For example, \beq
\hat{\psi}'_1\equiv\psi'_1-\left(\frac{n-1}{n}\right)\psi_0-\frac{1}{nm}
\not{\!\partial\!}\psi_0, \eeq{d2} is St\"uckelberg invariant,
because $\lambda'_1=\lambda_0$. Out of the invariant spinors
$\{\hat{\psi}'_1, \hat{\psi}'_2, ..., \hat{\psi}'_{n-1},
\hat{\psi}_n\}$ one can further construct another set of invariant
spinors $\{\tilde{\psi}'_1, \tilde{\psi}'_2, ...,
\tilde{\psi}'_{n-1}, \tilde{\psi}_n\}$, such that when the latter
set replaces its untwiddled counterpart in the gauge fixed
Lagrangian, it exactly reproduces the St\"uckelberg invariant
KK-reduced Lagrangian \emph{before} gauge fixing. The twiddled fields
are constructed by considering the gauge fixing conditions. Note
that $\tilde{\psi}_n$ and $\psi_n$ are related by the gauge
transformation: \beq \tilde{\psi}_n=\psi_n+\partial\lambda_{n-1}
=\psi_n+\partial\lambda_{n-1}^{\gamma T}+\frac{1}{4}\,\gamma
\partial\lambda_{n-2}.\eeq{d3} The gauge fixing conditions are such
that the highest power of $1/m$ in $\lambda_{k-1}$ is $(1/m)^k$.
Thus only the term $\partial\lambda_{n-1}^{\gamma T}$ in~(\ref{d3})
can contain $(1/m)^n$. In view of~(\ref{d1}) we have \beq
\tilde{\psi}_n=\psi_n+\frac{(-1)^n}{n!\,m^n}\left[\partial^n
+b_1(n)\gamma\partial^{n-1}\not{\!\partial\!}+b_2(n)\eta\partial^{n-2}
\Box +b_3(n)\eta\gamma\partial^{n-3}\Box\not{\!\partial\!}
+...\right]\psi_0+~...\eeq{d4} where $b(n)$'s are rational functions
of $n$ such that the quantity in the brackets is the derivative of a
gamma-traceless quantity (with indices symmetrized).

The St\"uckelberg invariant \emph{gauged} Lagrangian contains
interaction terms proportional to $e/m^{2n}=e/m^{2s-1}$. Such terms
come only from the kinetic pieces of the spin-$s$ field.
Furthermore, since the \emph{free} spin-$s$ kinetic operator acting
on $\partial\lambda_{n-1}^{\gamma T}$ gives zero, we only need to
take into account the terms that have $\mathcal{O}(e)$ contribution
from the kinetic operator, and $\mathcal{O}(1)$ contribution from
$\tilde{\psi}_n$. In other words, we need to consider the follwoing
\beq L_{\text{int}}=\bar{\tilde{\psi}}_{\mu_1 \mu_2...\mu_n}^{*(0)}
ieT^{\mu_1\mu_2...\mu_n;\nu_1\nu_2
...\nu_n}_1\tilde{\psi}_{\nu_1\nu_2...\nu_n}^{(0)}+L_{\leq(2s+2)}
+\mathcal{O}(e^2),\eeq{d5} where $\tilde{\psi}_{\mu_1\mu_2...
\mu_n}^{(0)}$ is given by Eq.~(\ref{d4}), and \bea T^{\mu_1\mu_2...
\mu_n;\nu_1\nu_2...\nu_n}_1&\equiv& \frac{n-1}{2}\,\eta^{\mu_3
\nu_3}... \eta^{\mu_n\nu_n} A_\rho\left\{(n/2)\,\eta^{\mu_1
\mu_2}\eta^{\rho\nu_1}+\eta^{\mu_1\nu_1}\gamma^{\mu_2\rho}
+\eta^{\mu_2\nu_1}\gamma^{\mu_1\rho}\right\}\gamma^{\nu_2}
\nonumber\\&&+~\frac{n-1}{2}\,\eta^{\mu_3\nu_3}...
\eta^{\mu_n\nu_n}A_\rho\left\{(\eta^{\mu_1\rho}\eta^{\nu_1\nu_2}
-\eta^{\mu_1\nu_1}\eta^{\rho\nu_2})\gamma^{\mu_2}+(\mu_1
\leftrightarrow\mu_2)\right\}\nonumber\\&&+~\frac{n}{4}\,
\eta^{\mu_3\nu_3}...\eta^{\mu_n\nu_n}A_\rho\left(\eta^{\mu_1\nu_2}
\gamma^{\mu_2\rho\nu_1}+\eta^{\mu_2\nu_2}\gamma^{\mu_1\rho\nu_1}\right)
\nonumber\\&&-~\frac{n-2}{4}\,\eta^{\mu_3\nu_3}...
\eta^{\mu_n\nu_n}A_\rho\left(\eta^{\mu_1\nu_1}\gamma^{\mu_2\rho\nu_2}
+\eta^{\mu_2\nu_1}\gamma^{\mu_1\rho\nu_2}\right)\nonumber\\&&+
~\frac{n(n-1)}{4}\,\eta^{\mu_3\nu_3}...\eta^{\mu_n\nu_n} A_\rho
\eta^{\mu_1\mu_2}\left(\eta^{\rho\nu_2}\gamma^{\nu_1} -\eta^{\nu_1
\nu_2}\gamma^\rho\right)\nonumber\\&&+~\frac{(n-1)(n-2)}{4}\,
\eta^{\mu_3\nu_3}...\eta^{\mu_n\nu_n} A_\rho\eta^{\nu_1\nu_2}
\left(\eta^{\rho\mu_1}\gamma^{\mu_2}+\eta^{\rho\mu_2}\gamma^{\mu_1}
\right), \eea{d6} with symmetrization assumed in ($\mu_1,\mu_2, ...,
\mu_n$) and in ($\nu_1,\nu_2, ..., \nu_n$). Given Eq.~(\ref{d5}), we
can easily compute the $\mathcal{O}(e)$ dimension-($2s+3$)
operators. We have \beq L_{2s+3}=-\frac{e}{2m^{2n}}\,
F_{\mu\nu}(\bar{\chi}
\overleftarrow{\partial}_{\mu_1}...\overleftarrow{\partial}_{\mu_{n-1}}
\gamma^{\mu\nu\rho}\partial_\rho\partial^{\mu_1}...\partial^{\mu_{n-1}}\chi)
~+~(...), \eeq{d7} where we denoted the spin-1/2 St\"uckelberg field
as $\psi_0\equiv\chi$, and the ellipses (...) stand for similar
terms containing $\not{\!\partial\!}\chi$. The terms in~(\ref{d7})
can be rendered proportional to the equations of motion, but the
degree of divergence cannot be improved by field redefinitions
followed by addition of local terms (similar to the cases of
spin-3/2 and -5/2). Addition of non-minimal terms does not help as
well. Therefore, in the scaling limit: $m\rightarrow0$ and
$e\rightarrow0$, such that $e/m^{2s-1}$=constant, the Lagrangian
becomes: \beq L=L_{\text{kin}}-\frac{e}{2m^{2n}}\,F_{\mu\nu}
(\bar{\chi}\overleftarrow{\partial}_{\mu_1}...\overleftarrow{\partial}_{
\mu_{n-1}}\gamma^{\mu\nu\rho}\partial_\rho\partial^{\mu_1}...\partial^{
\mu_{n-1}} \chi)~+~(...).\eeq{d8} The effective field theory has an
intrinsic finite cutoff \beq \Lambda_s=\frac{m}{e^{1/(2s-1)}}\,.
\eeq{d9}

\section{Conclusion}
In this paper we argued that massive, charged particles of spin
$s\geq 3/2$ {\em can} be described by a local effective action at
energy scales parametrically higher than their mass. This effective
action breaks down at or below an energy $E = {\cal O}(\Lambda_s)$,
$\Lambda_s= me^{-1/(2s-1)}$.

The cutoff $\Lambda_s$ can be improved neither by adding local
non-minimal coupling terms nor by local field redefinitions. For the
bosonic case, the latter result was proved using a cohomological
argument valid for all spins, described in details in Appendix A.
Its physical meaning is illustrated for the case of spin-2 particles
in Appendix B. The meaning of the cutoff $\Lambda_s$ is only that
some new physics must happen at a scale not higher than $\Lambda_s$. This
new physics could result in a strong coupling unitarization, or in
the existence of new interacting degrees of freedom, lighter than
the cutoff. In the first case, the theory becomes, to all effects,
non-local at a scale not higher than $\Lambda_s$, because the
spin-$s$ particle develops a form factor which implies a finite,
nonzero charge radius. In the second case, one could integrate out
the new light degrees of freedom, but the resulting action would be
non-local already below the scale $\Lambda_s$. The very possibility
of introducing non-local counter-terms into the action invalidates
the cohomological argument of Appendix A. This is the technical
reason why {\em lighter} degrees of freedom may be essential for a
complete UV embedding of our effective actions.

Notice that the examples given in the introduction exhibit both new
light degrees of freedom and non-localities at energy scales
parametrically smaller than $\Lambda_s$. High-spin hadronic
resonances have a natural inverse size ${\cal O}(\Lambda_{QCD})\ll
m$, and also interact with other lighter, lower-spin resonances. The
Argyres-Nappi action~\cite{an} describes a perturbative string
state, with intrinsic inverse size ${\cal O}(M_S) \ll m$, which
interacts with and can decay into many other lighter string states.

A third possibility, namely that the theory reaches a UV fixed
point, is not really different from strong coupling unitarization.
The point is that the intrinsic cutoff we found was due to the
explicit presence of power counting, non-renormalizable operators in
the effective action of the spin-$s$ particle. To make these
operators UV irrelevant, their anomalous dimensions must change
(run) by factors of order one from their tree-level values within a
small energy window ($m < E < \Lambda_s$). For instance, in the case
of spin-2, the operator~(\ref{r30}) must change its dimension from 7
at $E=m$ to 4 or less at $E=me^{-1/3}$! This cannot be achieved by
the logarithmic running implied by a perturbative effective local
action containing only a massive spin-2 field and the photon: a
dramatic, non-perturbative re-arrangement of the degrees of freedom
must take place {\em below the scale $\Lambda_s$}.

If a UV fixed point does exist, what would it look like? This is
difficult to say; what is easy to predict is that it will {\em not}
look like a weakly interacting spin-$s$ particle. This is because a
UV fixed point must necessarily describe a massless high-spin
particle. But, starting at spin-5/2, massless particles are
forbidden to interact with any lower-spin fields (including
graviton) by powerful no-go theorems~\cite{w64,ad,p}. For spin-2, a
weaker theorem~\cite{p} implies that the particle must be
electrically neutral, while all non-minimal terms, except for
possibly a dipole, are ruled out because they cannot have dimension
four, as demanded by the conformal invariance of the UV fixed point.

It is instructive to compare our results with claims contained in a
recent preprint~\cite{zinolast}\footnote{That preprint appeared
about a month later than our paper.}, where it is argued that, in
the case of massive, charged spin-2, a new choice of St\"uckelberg
fields may yield a higher cutoff than what we estimated in this
paper. This is of course impossible. First of all, the action
presented in~\cite{zinolast} is incomplete, since it does not
contain the EM kinetic term, whose gauge completion under its new
St\"uckelberg transformations contains non-renormalizable
interactions, singular in the massless limit. More importantly, {\em
any} gauge-invariant completion must reduce in the unitary gauge to
the minimal EM-coupled action plus a particular choice of
non-minimal terms. Otherwise, the action would not describe a spin-2
only, but also other {\em physical} degrees of freedom even in
perturbation theory. Once in the unitary gauge, we can use our
procedure to introduce {\em our} set of St\"uckelberg fields and
proceed to show that the cutoff is as in Eq.~(\ref{m1}). The core of
our paper is indeed proving that the cutoff~(\ref{m1}) cannot be
changed by any field redefinition or addition of non-minimal terms.
Moreover, the action considered in~\cite{zinolast} contains a vertex
$DhFDh$, needed to give a correct counting of degrees of freedom.
Although such a term is proportional to $e/m^2$, one \emph{cannot}
conclude that the model has a cutoff of $m/\sqrt{e}$. This is
because in the unitary gauge the propagator is singular in the
massless limit, so that one cannot simply read off the cutoff by
looking at the most divergent term. In other words, higher
dimensional operators have canonical dimensions \emph{only} when the
propagator has good high energy behavior (i.e. $\sim 1/p^2$). This
can be achieved by introducing St\"{u}ckelberg fields and performing
an appropriate covariant gauge-fixing. Only after the gauge fixing
can one correctly identify the canonical dimensions of the higher
dimensional operators. In particular, the true cutoff of the action
in~\cite{zinolast} is $me^{-1/6}$, which is much lower than our
upper bound of $me^{-1/3}$.

We remind again the reader that our aim was to find the maximum
cutoff of an effective Lagrangian of a single spin-$s$ field
interacting with EM. Our Lagrangian should be thought only as a
generating functional of the S-matrix. We are after a bound valid in
perturbation theory, and not after a Lagrangian free of pathologies,
which would require more powerful methods than the one we used; i.e.
methods that can fix terms quadratic and higher order in the EM
field. We emphasize that our theoretical maximum may never be
reached in a truly consistent theory. For instance, the spin-2
Argyres-Nappi Lagrangian~\cite{an} has a cutoff much lower than
$me^{-1/3}$.

We would like to conclude with a comment on spin-1 and an
observation on a possible new approach to the problem of finding the
intrinsic cutoff.

\subsubsection*{Spin-1}

The cutoff of a charged massive spin-1 particle is known to be
${\cal O}(m/e)$. In this special case we also know how to improve
the UV behavior of the theory: it suffices to add another degree of
freedom, a neutral Higgs scalar lighter than ${\cal O}(m/e)$. To
show that the cutoff in the absence of the neutral Higgs is no
higher than  ${\cal O}(m/e)$, we start by writing the Lagrangian of
a complex, massive spin-1 field $W_\mu$: \beq
L=-\frac{1}{2}\,|\partial_\mu W_\nu -\partial_\nu W_\mu|^2 -m^2
W_\mu^*W^\mu. \eeq{o1} We make it gauge invariant by introducing a
scalar (St\"uckelberg) field $\phi$ through the substitution $W_\mu
=V_\mu -\partial_\mu \phi/m$. The minimal substitution $\partial_\mu
\rightarrow D_\mu \equiv \partial_\mu + ieA_\mu$ preserves the
gauged St\"uckelberg symmetry: \bea \delta V_\mu
&=&D_{\mu}\lambda,\\ \delta\phi &=& m\lambda. \eea{o2} One thus
obtains the following non-renormalizable operators \beq
L_{\text{div}}=-\left[\frac{ie}{2m}\,F^{\mu\nu} \phi^*(D_\mu V_\nu
-D_\nu V_\mu) +\text{c.c.}\right]-
\frac{e^2}{2m^2}\,F_{\mu\nu}^2\phi^*\phi. \eeq{o3} The presence of
non-renormalizable interactions implies the existence of a UV cutoff
$\Lambda\sim m/e$. The tri-linear operators can be canceled by
adding to the Lagrangian the dipole term $ieF^{\mu\nu} W^*_\mu
W_\nu$. The resulting non-minimal Lagrangian still contains
tri-linear non-renormalizable operators: \beq
L_{\text{div}}^{(\text{NM})}=-\left[\frac{ie}{m}\,\partial_\mu
F^{\mu\nu} V^*_\nu \phi + \text{c.c.}\right] -\frac{ie}{m^2}\,
\partial_\mu F^{\mu\nu}\phi^*\partial_\nu\phi+{\cal O}(e^2). \eeq{o4}
Notice that the last term implies a cutoff $m/\sqrt{e}$,
parametrically lower than $m/e$! This hardly seems like a progress,
but since all non-renormalizable terms in Eq.~(\ref{o4}) are
proportional to the photon equations of motion, they can be canceled
by the local field redefinition: \beq A_\mu \rightarrow A_\mu +
\left\{\frac{ie}{m}\left(V^*_\mu \phi+\frac{1}{2m}\phi^*\partial_\mu
\phi\right)+\text{c.c.}\right\}. \eeq{o5} This redefinition
generates a host of dimension-6 and dimension-7 operators, of
schematic form $(e^2/m^2) O_6$, $(e^2/m^3) O_7$, and a dimension-8
operator proportional to $e^2/m^4$. The latter comes form the
redefinition of the Maxwell action. It is the most dangerous one,
since it introduces a lower cutoff $m/\sqrt{e}$; but it can be
eliminated by adding appropriate local counter-terms. All remaining
non-renormalizable terms are ${\cal O}(e^2)$, so that even though
they may not $-$ indeed cannot $-$ be canceled by adding {\em local}
counter-terms, they may be $-$ indeed are $-$ canceled by the {\em
non-local} counter-terms obtained by embedding the theory into a
spontaneously broken gauge theory with, e.g., gauge group $SU(2)$
broken to $U(1)$, and integrating out the neutral Higgs degree of
freedom.

Notice that in this case one obtains the same UV scale $m/e$ from
either the minimal Lagrangian or the non-minimal one. Indeed, unless
we knew in advance that a UV completion existed, the exercise of
adding the dipole term $ieF^{\mu\nu} W^*_\mu W_\nu$ would have made
the UV behavior of the theory {\em worse}, lowering the cutoff from
$m/e$ to $me^{-2/3}$!

\subsubsection*{Future Directions}
The explicit example of the Argyres-Nappi action~\cite{an}
potentially holds for us one last lesson. Its true cutoff is $M_S$,
much less than our ``optimal'' one, $\Lambda_2$. Even ignoring the
fact that this action is embedded in string theory, it still
contains a host of non-renormalizable operators proportional to
powers of $e F_{\mu\nu}/M_S^2$; therefore, already the interactions
involving only transverse modes become strong at the scale
$M_S/\sqrt{e}\ll \Lambda_2$. On the other hand, these non-minimal
terms are necessary to ensure that the theory propagates causally
five degrees of freedom in constant external EM fields~\cite{an}.
This observation opens up the possibility of obtaining more
stringent bounds on the UV cutoff of any high-spin theory by
requiring causality in external backgrounds. This is similar, at
least in spirit, to causality bounds found for certain effective
field theories in~\cite{aahdnr} and, in other theories and with
different methods, in~\cite{hm}.

\subsection*{Acknowledgments}
We would like to thank J. Polchinski, L. Rastelli, and J. Schwarz
for useful comments. MP is supported in part by NSF grants
PHY-0245068 and PHY-0758032. RR is partially supported by James
Arthur graduate fellowship.

\section*{Appendix A}
\setcounter{equation}{0}
\renewcommand{\theequation}{A.\arabic{equation}}

In this Appendix  we are going to use cohomological arguments to show
that for integer spin $s$ it is
impossible to eliminate up to a total derivative the dangerous
$\mathcal{O}(e^2)$ dimension-($4s+3$) operator by adding local counter-terms.
Note that the former
arises after we make a field redefinition of the $U(1)$ gauge field to
cancel the dimension-($2s+4$) operator, and then add a local term to
eliminate the resulting $\mathcal{O}(e^2)$ dimension-($4s+4$)
operator. We start with the simplest case of $s=2$. Then we use
similar arguments for spin-3, and finally generalize them for
arbitrary spin $s$.

For spin-2, we want to study the cohomology defined by the
coboundary operator: \beq \delta = \int d^4 x \left[(c+ c_\mu x^\mu
){\delta \over \delta \phi (x)} + (b_\mu + b_{\mu\nu}x^\nu) {\delta
\over \delta B_\mu (x)}\right], \eeq{coho1} on space-time integrals
of local functionals of the basic fields $\phi(x)$, $B_\mu(x)$, and
derivatives thereof, modulo total derivatives. That is, we want to
find the space of functionals $P(\phi, B_\mu, \partial_\nu\phi,...)$
which obey $\delta \int d^4 x P = 0 $, and are not of the form $\int
d^4 x P = \delta \int d^4 x P'$ for some other $P'(\phi, B_\mu,
\partial_\nu\phi,...)$. Here $c,c_\mu,b_\mu,b_{\mu\nu}$ are
space-time constant anti-commuting variables. We allow no explicit
dependence on the coordinates $x^\mu$ in either $P$ or $P'$. The
integrand $P$ is defined modulo a total derivative, of course. To
take this fact into account, we define another coboundary operator:
$d\equiv\epsilon^\mu\partial_\mu$, with $\epsilon^\mu$ an
anti-commuting variable. Prior to defining a scalar product and a
Hermitian conjugate for $d$, $\epsilon^\mu$ is just another name for
the coordinate differential $dx^\mu$. With the help of $d$, the
cohomology now reads \beq \delta P = d Q, \qquad P \neq \delta P' +
d Q'. \eeq{coho2} To study the cohomology it is extremely useful to
define a Hilbert space structure for the space of $P$'s. It allows
us to define Hermitian conjugates for $\delta$ and $d$.

A Hilbert space structure can be defined on local functionals of
fields as follows~\cite{dix}. We Taylor expand a field $A(x)$ around
$x^\mu =0$: \beq A(x)= \sum_{n=0}^\infty {1\over n!}
a^*_{\mu_1...\mu_n}x^{\mu_1}...x^{\mu_n}. \eeq{coho3} Then we
promote each $a_{\mu_1...\mu_n}^*$ to a canonical Bosonic creation
operator; the corresponding canonical annihilation operator
$a_{\mu_1...\mu_n}$ is its Hermitian conjugate. By definition, we
have the commutation relations: \beq [a_{\mu_1...\mu_n},
a^{*\,\nu_1...\nu_n}]=\sum \delta_{\mu_1}^{\nu_{p(1)}}...
\delta_{\mu_n}^{\nu_{p(n)}}, \eeq{coho4} where the sum extends to
all permutations $p(1),..., p(n)$ of $n$ indices. Likewise,
Hermitian conjugates of $c,c_\mu,b_\mu,b_{\mu\nu},\epsilon^\mu$ are
defined by imposing canonical anti-commutation relations for all
these variables. In particular: \beq \{ \epsilon^\mu,
\epsilon^*_\nu\}= \delta^\mu_\nu. \eeq{coho5} A local function of
the field $A$ and its derivatives defines a vector in the Fock
space, which is obtained by applying the creation operators
$a^*_{\mu_1 ... \mu_n}$ to the vacuum state $|0\rangle $. We have
$a_{\mu_1...\mu_n}|0\rangle=0$ for all $a$'s. The correspondence is
\beq P(A, \partial_\mu A,....) \rightarrow P(a^*, a^*_\mu,....)
|0\rangle. \eeq{coho6} On the Fock space, the derivative operator
$\partial_\mu$ reduces to~\cite{dix}\footnote{Opposite
to~\cite{dix}, we use the standard physicists' conventions for
creation and annihilation operators.} \beq
\partial_\mu= \sum_{n=0}^\infty {1\over n!}
a^*_{\mu_1 ... \mu_n\mu}a^{\mu_1 ... \mu_n}. \eeq{coho7} Its
Hermitian conjugate is therefore \beq
\partial^*_\mu= \sum_{n=0}^\infty {1\over n!}
a^*_{\mu_1 ... \mu_n}a_\mu^{~\mu_1 ... \mu_n}. \eeq{coho8} Thanks to
these definitions and to the anti-commutation
relations~(\ref{coho5}), the Laplacian operator $\Delta\equiv (d +
d^*)^2$ can be written as \beq \Delta = \epsilon^{*\,\mu}
\partial_\mu^* \epsilon^\nu\partial_\nu + \epsilon^\nu
\partial_\nu\epsilon^{*\,\mu}\partial_\mu^* = \partial^\mu
\partial_\mu^* + \epsilon^*_{\mu}\epsilon^\mu N, \qquad
N\equiv \sum_{n=0}^\infty {1\over n!} a^*_{\mu_1 ... \mu_n}a^{\mu_1
... \mu_n}. \eeq{coho9} On a monomial in the field and its
derivatives, the particle number operator $N$ reduces to the degree
of the monomial. Eq.~(\ref{coho9}) uses the identity~\cite{dix}:
\beq [\partial^*_\mu,\partial^\nu]=\delta_\mu^\nu N, \eeq{coho10}
which in turn can be easily derived using definitions~(\ref{coho7},
\ref{coho8}), and the commutator \bea [a^*_{\mu_1 ... \mu_n}a^{\mu_1
... \mu_n\mu}, a^*_{\nu_1 ... \nu_n\nu}a^{\nu_1 ...
\nu_n}]&=&{1\over n!} \delta^\mu_\nu a^*_{\mu_1 ... \mu_n}a^{\mu_1
... \mu_n} + {1\over (n-1)!} a^*_{\mu_1 ... \mu_{n-1}\nu}a^{\mu_1
... \mu_{n-1}\mu} \nonumber
\\ && -{1\over n!} a^*_{\mu_1 ... \mu_{n}\nu}a^{\mu_1 ... \mu_{n}\mu}.
\eea{coho11} The identification $\epsilon^\mu = dx^\mu$ implies that
$\epsilon_\mu^* \epsilon^\mu=4-D_F$, where $D_F$ counts the degree
of the differential form obtained by the replacement $\epsilon^\mu
\rightarrow dx^\mu$, $a_{\mu_1 ... \mu_n} \rightarrow
\partial_{\mu_1}... \partial_{\mu_n}A$ in the Hilbert space vector
$P(\epsilon,a,a_\mu,...)|0\rangle$.

Now the $d$ cohomology is isomorphic to the space of harmonic
vectors, which are solutions of the equation $\Delta \psi=0$. Since
$\Delta= (4-D_F)N$ + positive operator, the cohomology is trivial
unless either $D_F=4$ or $N=0$. This fact is useful for our purpose.

Let us now apply this machinery to our problem, where the fields are
$\phi$ and $B_\mu$. For spin-2, the dimension-11 operator we want to
cancel by adding local counter-terms is \beq
L_{11}=\frac{e^2}{2m^7}\,\{\partial^\mu\partial_\sigma\phi^*
\partial^\nu\partial^\sigma\phi-(\mu\leftrightarrow\nu)\}\,\{2\partial_{[\mu}
B^*_{\rho]}\partial^\rho\partial_\nu\phi\ - \partial_{[\mu}
B_{\nu]}^*\Box \phi\}. \eeq{coho12} We can write it as a four-form
with $N=4$, denoted by $O_{(4,4)}$: \bea O_{(4,4)}=
\frac{1}{4!}\,\varepsilon_{\alpha\beta\gamma\delta}L_{11}\epsilon^{\alpha}
\epsilon^{\beta}\epsilon^{\gamma}\epsilon^{\delta}.\eea{coho13} Its
variation under the operator $\delta$, defined in Eq.~(\ref{coho1}),
is a nonzero total derivative: a three form with $N=3$. By labeling
all operators appearing in our cohomology equations as $O_{(D_F,N)}$
we get the equation \beq \delta O_{(4,4)}= d O_{(3,3)}. \eeq{coho14}
By applying $\delta$ to both sides of this equation and using the
fact that the $d$ cohomology is soluble on operators with $N=2,
D_F=3$ we get \beq \delta O_{(3,3)}= d O_{(2,2)}. \eeq{coho15}
Repeating the same argument on $O_{(2,2)}$ we get $\delta O_{(2,2)}=
d O_{(1,1)}$ and finally \beq d\delta O_{(1,1)}=0. \eeq{coho16} Now,
$\delta O_{(1,1)}$ is a one form with $N=0$, thus it also obeys
$\partial^*_\mu \delta O_{(1,1)}=0$. So, unless it vanishes
identically, it is a nontrivial element of the $d$ cohomology. In
particular, \beq \delta O_{(1,1)}\neq d O_{(0,0)}. \eeq{coho17} In
the spin-2 case, an explicit computation gives\footnote{To avoid
confusion, let us point out that $\bar{c}$, $\bar{b}_\mu$ etc. are
the ghost fields associated with the complex conjugate fields
$\phi^*$ and $B^*_\mu$, {\em not} the Hermitian conjugates of $c$,
$b_\mu$ etc.!} \beq \delta O_{(1,1)}=\frac{6e^2}{m^7}\,
\varepsilon^{\mu\nu\rho\sigma}\bar{c}_\alpha c^\alpha
\bar{b}_{\mu\nu} c_\rho\epsilon_\sigma \neq 0. \eeq{coho18}

Now, in order to cancel the dangerous dimension-11 operator
$O_{(4,4)}$, there must exist another four-form $\tilde{O}_{(4,4)}$,
of the same dimension, such that the sum of the two is a total
derivative \beq O_{(4,4)}+\tilde{O}_{(4,4)}= d O_{(3,4)}.
\eeq{coho19} The operator $\tilde{O}_{(4,4)}$ is built with the
St\"uckelberg fields, so that by construction it is annihilated by
the operator $\delta$; this gives us the equation \beq \delta
O_{(4,4)}= d \delta O_{(3,4)}. \eeq{coho20} Comparing
Eqs.~(\ref{coho20}) and~(\ref{coho14}) we find that $\delta
O_{(3,4)} - O_{(3,3)}$ is closed. Since the $d$ cohomology is
trivial on three-forms with $N=3$ we get \beq O_{(3,3)}= \delta
O_{(3,4)} + d O_{(2,3)}. \eeq{coho21} By applying $\delta$ to both
sides of Eq.~(\ref{coho21}), comparing the result with
Eq.~(\ref{coho15}), and using the triviality of the $d$ cohomology
on two-forms with $N=2$, and then repeating the argument once more,
on one forms with $N=1$, we finally arrive at \beq O_{(1,1)}= \delta
O_{(1,2)} + d O_{(0,1)}. \eeq{coho22} This immediately says that
$\delta O_{(1,1)}$ is $d$-exact, in contradiction with
Eq.~(\ref{coho17}). The conclusion is that the dangerous
dimension-11 operator cannot be eliminated, up to a total
derivative, by adding local counter-terms.

The analysis for spin-3 is quite similar. In view of
Eqs.~(\ref{t32}, \ref{t33}), the coboundary operator $\delta$ is
given by \beq \delta = \int d^4 x \left[\left(c+c_\mu
x^\mu+\frac{1}{2}\,c_{\mu\nu} x^\mu x^\nu\right){\delta \over \delta
\phi (x)} + \left(b_\mu + b_{\mu\nu}x^\nu+\frac{1}{2}\,
b_{\mu\nu\rho}x^\nu x^\rho \right) {\delta \over \delta B_\mu
(x)}\right]. \eeq{coho23} The four-form under consideration is \bea
O_{(4,4)}=\frac{1}{4!}\, \varepsilon_{\alpha
\beta\gamma\delta}L_{15}\epsilon^{\alpha}
\epsilon^{\beta}\epsilon^{\gamma}\epsilon^{\delta}, \eea{coho24}
where $L_{15}$ is the dimension-15 operator~(\ref{t31}) we want to
eliminate. Here again, we have a nonzero $\delta O_{(1,1)}$: \beq
\delta O_{(1,1)}=\frac{6e^2}{m^{11}}\,\varepsilon^{\mu\nu\rho\sigma}
\bar{c}_{\alpha\beta} c^{\alpha\beta} \bar{b}_{\mu\nu\lambda}
c^\lambda_{~\rho}\epsilon_\sigma \neq 0. \eeq{coho25} This leads us
into contradictions if we want to eliminate~(\ref{coho24}) by local
counter-terms.

We expect a similar story for generic spin $s$, where the operator
$\delta$ is defined in accordance with Eqs.~(\ref{b14}, \ref{b15}). After all,
the only difference with the previous examples is a much larger number of
Lorentz indices, which we expect to entail algebraic complications but no
qualitatively new feature.
Once again, we expect a nonzero $\delta O_{(1,1)}$, which would make it
impossible to eliminate the $\mathcal{O}(e^2)$
dimension-($4s+3$) operator.

\section*{Appendix B}
\setcounter{equation}{0}
\renewcommand{\theequation}{B.\arabic{equation}}

In this Appendix, we give a physical explanation of the fact that for
spin-2 the dimension-11 operator~(\ref{r25}) cannot be eliminated,
up to a total derivative, by adding local counter-terms. We have
\beq L_{11}=\frac{e^2}{2m^7}\,\{\partial^\mu\partial_\sigma\phi^*
\partial^\nu\partial^\sigma\phi-(\mu\leftrightarrow\nu)\}\,\{2
\partial_{[\mu}B^*_{\rho]}\partial^\rho\partial_\nu\phi\
-\partial_{[\mu} B_{\nu]}^*\Box \phi\}. \eeq{sc1} Let us consider an
explicit scattering process involving two $\phi$'s, one $\phi^*$,
and one $B_\mu^*$. In the process under consideration, the $\phi$'s
and $\phi^*$ are incoming and on-shell, which produce an off-shell
$B_\mu^*$ at rest. In the rest frame of $B_\mu^*$, let the 4-momenta
of the two $\phi$'s be \beq k_1^\mu= E(1, \cos\theta, \sin\theta,
0), \qquad k_2^\mu= E(1, \cos\theta, -\sin\theta, 0), \eeq{sc2} and
that of the $\phi^*$ be \beq p^\mu=2E(\cos\theta, -\cos\theta, 0,
0). \eeq{sc3} We have $k_1^2=k_2^2=p^2=0$, which are just on-shell
conditions in the limit $m\rightarrow0$. The following
scalar products will be useful: \bea k_1\cdot k_2&=&-2E^2\sin^2\theta,\label{sc4}\\
p\cdot k_1&=&-2E^2 \cos\theta(1+\cos\theta)~=~p\cdot k_2~.
\eea{sc5} We define
$G_{\mu\nu}\equiv\partial_{[\mu}B_{\nu]}$, and
$H_{\mu\nu}\equiv\partial_{(\mu}B_{\nu)}$. The momentum space
amplitude that comes from~(\ref{sc1}) is given by\footnote{In this Appendix
we set $e^2/m^7=1$.} \bea \mathcal{A}&=&-G^*_{\mu\nu}
(k_1\cdot k_2)\{(p\cdot k_1)p^\mu k_2^\nu+(p\cdot k_2)p^\mu k_1^\nu\}\nonumber\\
\nonumber\\&=&-16E^6\sin^2\theta\cos^2\theta(1+\cos\theta)^2
G^*_{01}. \eea{sc6}

We would like to cancel this amplitude by adding some local
counter-terms in the Lagrangian. Let us add to the Lagrangian the
most general set of counter-terms of appropriate dimension: \bea
L_{\text{c-t}}&=&e^2\,[\,\alpha\,\text{Tr}(\tilde{h}\tilde{h}^*
\tilde{h}\tilde{h}^*)+\beta\,\text{Tr}(\tilde{h}\tilde{h}\tilde{h}^*
\tilde{h}^*)+\gamma\,\text{Tr}(\tilde{h}^*\tilde{h}\tilde{h}^*)\,
\text{Tr}\tilde{h}\nonumber\\&&~~~+\gamma^*\,\text{Tr}
(\tilde{h}\tilde{h}^*\tilde{h})\,\text{Tr}\tilde{h}^*+\delta\,\text{Tr}
(\tilde{h}\tilde{h}^*)\,\text{Tr}(\tilde{h}\tilde{h}^*)+\varepsilon\,
\text{Tr}(\tilde{h}\tilde{h})\,\text{Tr}(\tilde{h}^*\tilde{h}^*)\,],
\eea{sc7} where $\alpha, \beta, \gamma, \delta,
\varepsilon$ are dimensionless constants. These counter-terms in
turn give new dimension-11 operators, some of which contribute to
the process under consideration. The contribution to the amplitude
coming from~(\ref{sc7}) is \bea \mathcal{A}'&=&2\beta
H^*_{\mu\nu}(k_1\cdot k_2)\{(p\cdot k_2) k_1^\mu p^\nu +(p\cdot
k_1)k_2^\mu p^\nu\}+2\delta H^*_{\mu\nu} \{(p\cdot k_1)^2k_2^\mu
k_2^\nu +(p\cdot k_2)^2k_1^\mu
k_1^\nu\}\nonumber\\&&+\{4\alpha H^*_{\mu\nu} k_1^\mu
k_2^\nu+2\gamma^* H^{*\mu}_\mu(k_1\cdot k_2)\} (p\cdot k_1) (p\cdot
k_2)+4\varepsilon H^*_{\mu\nu}(k_1\cdot k_2)^2 p^\mu
p^\nu.\eea{sc8}  We want to have $\mathcal{A}+
\mathcal{A}'=0$, for any possible configuration of $B^*_\mu$, and
for any angle $\theta$.

Let us choose the configuration: $B_\mu= (B_0(t), 0, 0, 0)$, for
which we have $G^*_{\mu\nu}=0$, and $H^*_{\mu\nu}=
\text{diag}(2\dot{B}^*_0, 0, 0, 0)$. In this case, $\mathcal{A}$
vanishes, and $\mathcal{A}'$ reduces to \beq \mathcal{A}'=
32\dot{B}^*_0 E^6\cos^2\theta[\,(1+\cos\theta)^2\{\alpha
+\delta+\gamma^* \sin^2\theta\}+2\sin^2\theta
\{\beta(1+\cos\theta)+2\varepsilon \sin^2\theta\}\,]. \eeq{sc9} This
must vanish for any $\theta$, in particular for $\theta=0$. This
gives $\delta=-\alpha$, so that we have \beq \mathcal{A}'=
32\dot{B}^*_0 E^6\cos^2\theta\sin^2\theta (1+\cos\theta)
[\,\gamma^*(1+\cos\theta)+2\beta+4\varepsilon (1-\cos\theta)\,].
\eeq{sc10} In order for this to vanish for generic $\theta$, we must
have $\gamma^*=-\beta$, $\varepsilon=-\beta/4$. All these conditions
constrain the amplitude~(\ref{sc8}) for any generic configuration of
$B^*_\mu$ to \bea \mathcal{A}'&=&-2\alpha H^*_{\mu\nu} (p\cdot
k_1)^2(k_1-k_2)^\mu(k_1-k_2)^\nu-2\beta H^{*\mu}_\mu(k_1\cdot
k_2)(p\cdot k_1)^2\nonumber\\&&+\beta H^*_{\mu\nu}
(k_1\cdot k_2)(p\cdot k_1)\{2(k_1+k_2)^\mu p^\nu-(k_1\cdot k_2)
p^\mu p^\nu\}.\eea{sc11}

Now we choose another configuration: $B_\mu= (0, B_1(t), 0, 0)$.
Then all components of $H^*_{\mu\nu}$, but $H^*_{01}=H^*_{10}$, are
zero, so that $H^{*\mu}_\mu=0$. Given $(k_1-k_2)^\mu=2E(0, 0,
\sin\theta, 0)$, and $(k_1+k_2)^\mu=2E(1, \cos\theta, 0, 0)$, one
finds that $\mathcal{A}'$ vanishes identically. Thus the two
amplitudes cannot cancel in general, as $\mathcal{A}$ does not
vanish in this case: \beq \mathcal{A}=-16E^6\sin^2\theta
\cos^2\theta(1+\cos\theta)^2 \dot{B}^*_1 \neq 0. \eeq{sc12} In
conclusion, no set of local counter-terms can eliminate $L_{11}$.
This signals a bad UV behavior, since
the amplitude goes like $E^6$, and therefore
blows up at high energies.


\begin{thebibliography}{99}
\bibitem{ad}
  C.~Aragone and S.~Deser,
  %``Consistency Problems Of Hypergravity,''
  Phys.\ Lett.\  B {\bf 86}, 161 (1979).
  %%CITATION = PHLTA,B86,161;%%
\bibitem{ww}
  S.~Weinberg and E.~Witten,
  %``Limits On Massless Particles,''
  Phys.\ Lett.\  B {\bf 96}, 59 (1980).
  %%CITATION = PHLTA,B96,59;%%
\bibitem{p}
M.~Porrati,
  %``Universal Limits on Massless High-Spin Particles,''
Phys. \ Rev. \ D {\bf 78}, 065016 (2008)
  [arXiv:0804.4672 [hep-th]].
  %%CITATION = ARXIV:0804.4672;%%
\bibitem{an}
  P.~C.~Argyres and C.~R.~Nappi,
  %``MASSIVE SPIN-2 BOSONIC STRING STATES IN AN ELECTROMAGNETIC BACKGROUND,''
  Phys.\ Lett.\  B {\bf 224}, 89 (1989).
  %%CITATION = PHLTA,B224,89;%%
\bibitem{vz}
  G.~Velo and D.~Zwanziger,
  %``Propagation And Quantization Of Rarita-Schwinger Waves In An External
  %Electromagnetic Potential,''
  Phys.\ Rev.\  {\bf 186}, 1337 (1969);
  %%CITATION = PHRVA,186,1337;%%
  %``Noncausality and other defects of interaction lagrangians for particles
  %with spin one and higher,''
  Phys.\ Rev.\  {\bf 188}, 2218 (1969).
  %%CITATION = PHRVA,188,2218;%%
  G.~Velo,
  %``Anomalous behaviour of a massive spin two charged particle in an external
  %electromagnetic field,''
  Nucl.\ Phys.\  B {\bf 43}, 389 (1972).
  %%CITATION = NUPHA,B43,389;%%
\bibitem{sh}
  L.~P.~S.~Singh and C.~R.~Hagen,
  %``Lagrangian formulation for arbitrary spin. 1. The boson case,''
  Phys.\ Rev.\  D {\bf 9}, 898 (1974);
  %%CITATION = PHRVA,D9,898;%%
  %``Lagrangian formulation for arbitrary spin. 2. The fermion case,''
  Phys.\ Rev.\  D {\bf 9}, 910 (1974).
  %%CITATION = PHRVA,D9,910;%%
\bibitem{z1}
Yu.~M.~Zinoviev,
  %``On massive high spin particles in (A)dS,''
  arXiv:hep-th/0108192;
  %%CITATION = HEP-TH/0108192;%%
  %``On massive mixed symmetry tensor fields in Minkowski space and (A)dS,''
  arXiv:hep-th/0211233.
  %%CITATION = HEP-TH/0211233;%%
\bibitem{z2}
  S.~M.~Klishevich and Yu.~M.~Zinovev,
  %``Electromagnetic interaction of a massive spin-2 particle,''
  Phys.\ Atom.\ Nucl.\  {\bf 61}, 1527 (1998)
  [Yad.\ Fiz.\  {\bf 61}, 1638 (1998)]
  [arXiv:hep-th/9708150].
  %%CITATION = YAFIA,61,1638;%%
\bibitem{met}
  R.~R.~Metsaev,
  %``Gauge invariant formulation of massive totally symmetric fermionic fields
  %in (A)dS space,''
  Phys.\ Lett.\  B {\bf 643}, 205 (2006)
  [arXiv:hep-th/0609029].
  %%CITATION = PHLTA,B643,205;%%
\bibitem{med}
  P.~de Medeiros,
  %``Massive gauge-invariant field theories on spaces of constant curvature,''
  Class.\ Quant.\ Grav.\  {\bf 21}, 2571 (2004)
  [arXiv:hep-th/0311254].
  %%CITATION = CQGRD,21,2571;%%
\bibitem{bian}
  M.~Bianchi, P.~J.~Heslop and F.~Riccioni,
  %``More on la grande bouffe,''
  JHEP {\bf 0508}, 088 (2005)
  [arXiv:hep-th/0504156].
  %%CITATION = JHEPA,0508,088;%%
\bibitem{hall}
  K.~Hallowell and A.~Waldron,
  %``Constant curvature algebras and higher spin action generating  functions,''
  Nucl.\ Phys.\  B {\bf 724}, 453 (2005)
  [arXiv:hep-th/0505255].
  %%CITATION = NUPHA,B724,453;%%
\bibitem{brst}
  I.~L.~Buchbinder and V.~A.~Krykhtin,
  %``Gauge invariant Lagrangian construction for massive bosonic higher spin
  %fields in D dimensions,''
  Nucl.\ Phys.\  B {\bf 727}, 537 (2005)
  [arXiv:hep-th/0505092].
  %%CITATION = NUPHA,B727,537;%%
  I.~L.~Buchbinder, V.~A.~Krykhtin, L.~L.~Ryskina and H.~Takata,
  %``Gauge invariant Lagrangian construction for massive higher spin fermionic
  %fields,''
  Phys.\ Lett.\  B {\bf 641}, 386 (2006)
  [arXiv:hep-th/0603212].
  %%CITATION = PHLTA,B641,386;%%
  I.~L.~Buchbinder, V.~A.~Krykhtin and P.~M.~Lavrov,
  %``\hfill{\normalsize{}hep-th/0608005,''
  Nucl.\ Phys.\  B {\bf 762}, 344 (2007)
  [arXiv:hep-th/0608005].
  %%CITATION = NUPHA,B762,344;%%
  I.~L.~Buchbinder, V.~A.~Krykhtin and A.~A.~Reshetnyak,
  %``BRST approach to Lagrangian construction for fermionic higher spin   fields
  %in AdS space,''
  Nucl.\ Phys.\  B {\bf 787}, 211 (2007)
  [arXiv:hep-th/0703049].
  %%CITATION = NUPHA,B787,211;%%
  I.~L.~Buchbinder, V.~A.~Krykhtin and H.~Takata,
  %``Gauge invariant Lagrangian construction for massive bosonic mixed symmetry
  %higher spin fields,''
  Phys.\ Lett.\  B {\bf 656}, 253 (2007)
  [arXiv:0707.2181 [hep-th]].
  %%CITATION = PHLTA,B656,253;%%
  P.~Y.~Moshin and A.~A.~Reshetnyak,
  %``BRST approach to Lagrangian formulation for mixed-symmetry fermionic
  %higher-spin fields,''
  JHEP {\bf 0710}, 040 (2007)
  [arXiv:0707.0386 [hep-th]].
  %%CITATION = JHEPA,0710,040;%%
\bibitem{frame}
  Yu.~M.~Zinoviev,
  %``Frame-like gauge invariant formulation for massive high spin particles,''
  Nucl.\ Phys.\  B {\bf 808}, 185 (2009)
  [arXiv:0808.1778 [hep-th]];
  %%CITATION = NUPHA,B808,185;%%
  %``Toward frame-like gauge invariant formulation for massive mixed symmetry
  %bosonic fields,''
  arXiv:0809.3287 [hep-th].
  %%CITATION = ARXIV:0809.3287;%%
\bibitem{bg}
  I.~L.~Buchbinder and A.~V.~Galajinsky,
  %``Quartet unconstrained formulation for massive higher spin fields,''
  JHEP {\bf 0811}, 081 (2008)
  [arXiv:0810.2852 [hep-th]].
  %%CITATION = JHEPA,0811,081;%%
\bibitem{partial}
  S.~Deser and A.~Waldron,
  %``Gauge invariances and phases of massive higher spins in (A)dS,''
  Phys.\ Rev.\ Lett.\  {\bf 87}, 031601 (2001)
  [arXiv:hep-th/0102166];
  %%CITATION = PRLTA,87,031601;%%
  %``Partial masslessness of higher spins in (A)dS,''
  Nucl.\ Phys.\  B {\bf 607}, 577 (2001)
  [arXiv:hep-th/0103198];
  %%CITATION = NUPHA,B607,577;%%
  %``Null propagation of partially massless higher spins in (A)dS and
  %cosmological constant speculations,''
  Phys.\ Lett.\  B {\bf 513}, 137 (2001)
  [arXiv:hep-th/0105181].
  %%CITATION = PHLTA,B513,137;%%
  E.~D.~Skvortsov and M.~A.~Vasiliev,
  %``Geometric formulation for partially massless fields,''
  Nucl.\ Phys.\  B {\bf 756}, 117 (2006)
  [arXiv:hep-th/0601095].
  %%CITATION = NUPHA,B756,117;%%
\bibitem{ff}
  C.~Fronsdal,
  %``Massless Fields With Integer Spin,''
  Phys.\ Rev.\  D {\bf 18}, 3624 (1978);
  %%CITATION = PHRVA,D18,3624;%%
  J.~Fang and C.~Fronsdal,
  %``Massless Fields With Half Integral Spin,''
  Phys.\ Rev.\  D {\bf 18}, 3630 (1978).
  %%CITATION = PHRVA,D18,3630;%%
\bibitem{ady}
  C.~Aragone, S.~Deser and Z.~Yang,
  %``MASSIVE HIGHER SPIN FROM DIMENSIONAL REDUCTION OF GAUGE FIELDS,''
  Annals Phys.\  {\bf 179}, 76 (1987).
  %%CITATION = APNYA,179,76;%%
\bibitem{rs}
  S.~D.~Rindani and M.~Sivakumar,
  %``Gauge - Invariant Description Of Massive Higher - Spin Particles By
  %Dimensional Reduction,''
  Phys.\ Rev.\  D {\bf 32}, 3238 (1985);
  %%CITATION = PHRVA,D32,3238;%%
  S.~D.~Rindani, D.~Sahdev and M.~Sivakumar,
  %``Dimensional reduction of symmetric higher spin actions. 1. Bosons,''
  Mod.\ Phys.\ Lett.\  A {\bf 4}, 265 (1989);
  %%CITATION = MPLAE,A4,265;%%
  %``Dimensional Reduction Of Symmetric Higher Spin Actions. 2: Fermions,''
  Mod.\ Phys.\ Lett.\  A {\bf 4}, 275 (1989).
  %%CITATION = MPLAE,A4,275;%%
\bibitem{pr1}
  M.~Porrati and R.~Rahman,
  %``Intrinsic Cutoff and Acausality for Massive Spin 2 Fields Coupled to
  %Electromagnetism,''
  Nucl.\ Phys.\  B {\bf 801}, 174 (2008)
  [arXiv:0801.2581 [hep-th]].
  %%CITATION = NUPHA,B801,174;%%
\bibitem{pr2}
  M.~Porrati and R.~Rahman,
  %``Electromagnetically Interacting Massive Spin-2 Field: Intrinsic Cutoff and
  %Pathologies in External Fields,''
  arXiv:0809.2807 [hep-th].
  %%CITATION = ARXIV:0809.2807;%%
\bibitem{fksc}
  P. Federbush,
  %``Minimal Electromagnetic Coupling for Spin Two Particles,''
  Nuovo Cimento {\bf 19}, 572 (1961).
  %%CITATION = PHRVA,186,1337;%%
  M.~Kobayashi and A.~Shamaly,
  %``Minimal Electromagnetic Coupling For Massive Spin-2 Fields,''
  Phys.\ Rev.\  D {\bf 17}, 2179 (1978);
  %%CITATION = PHRVA,D17,2179;%%
  %``The Tenth Constraint In The Minimally Coupled Spin-2 Wave Equations,''
  Prog.\ Theor.\ Phys.\  {\bf 61}, 656 (1979).
  %%CITATION = PTPKA,61,656;%%
  A.~Shamaly and A.~Z.~Capri,
  %``Propagation Of Interacting Fields,''
  Annals Phys.\  {\bf 74}, 503 (1972).
  %%CITATION = APNYA,74,503;%%
\bibitem{z3}
  Yu.~M.~Zinoviev,
  %``On massive spin 2 interactions,''
  Nucl.\ Phys.\  B {\bf 770}, 83 (2007)
  [arXiv:hep-th/0609170];
  %%CITATION = NUPHA,B770,83;%%
\bibitem{z4}
  Yu.~M.~Zinoviev,
  %``On spin 2 electromagnetic interactions,''
  arXiv:0806.4030 [hep-th].
  %%CITATION = ARXIV:0806.4030;%%
\bibitem{d2}
  S.~Deser,
  %``The Limit Of Massive Electrodynamics,''
  Annales Poincare Phys.\ Theor.\  {\bf 16}, 79 (1972).
  %%CITATION = AHPAA,16,79;%%
  S.~Deser and A.~Waldron,
  %``Inconsistencies of massive charged gravitating higher spins,''
  Nucl.\ Phys.\  B {\bf 631}, 369 (2002)
  [arXiv:hep-th/0112182].
  %%CITATION = NUPHA,B631,369;%%
  %``Acausality of massive charged spin 2 fields,''
  arXiv:hep-th/0304050.
  %%CITATION = HEP-TH/0304050;%%
  %``Partially massless spin 2 electrodynamics,''
  Phys.\ Rev.\  D {\bf 74}, 084036 (2006)
  [arXiv:hep-th/0609113].
  %%CITATION = PHRVA,D74,084036;%%
\bibitem{pfn}
  M.~Fierz and W.~Pauli,
  %``On relativistic wave equations for particles of arbitrary spin in an
  %electromagnetic field,''
  Proc.\ Roy.\ Soc.\ Lond.\  A {\bf 173}, 211 (1939);
  %%CITATION = PRSLA,A173,211;%%
  %``ON RELATIVISTIC FIELD EQUATIONS OF PARTICLES WITH ARBITRARY SPIN IN AN
  %ELECTROMAGNETIC FIELD,''
  Helv.\ Phys.\ Acta {\bf 12}, 297 (1939).
  %%CITATION = HPACA,12,297;%%
  P.~Van Nieuwenhuizen,
  %``On Ghost-Free Tensor Lagrangians And Linearized Gravitation,''
  Nucl.\ Phys.\  B {\bf 60}, 478 (1973).
  %%CITATION = NUPHA,B60,478;%%
\bibitem{kliboson}
  S.~M.~Klishevich,
  %``Massive fields of arbitrary half-integer spin in constant  electromagnetic
  %field,''
  Int.\ J.\ Mod.\ Phys.\  A {\bf 15}, 609 (2000)
  [arXiv:hep-th/9811030].
  %%CITATION = IMPAE,A15,609;%%
\bibitem{lagspin3}
  C.~C.~Chiang,
  %``Lagrangian formalism for neutral, massive spin 3 fields,''
  Prog.\ Theor.\ Phys.\  {\bf 45}, 1311 (1971).
  %%CITATION = PTPKA,45,1311;%%
  M.~Kawasaki, M.~Kobayashi and Y.~Mori,
  %``Lagrange Formulation Of The 20 Component Theory Of Spin 3 Fields,''
  Lett.\ Nuovo Cim.\  {\bf 14}, 611 (1975).
  %%CITATION = NCLTA,14,611;%%
  S.~C.~Lim,
  %``Lagrangian Formulation Of Massive Spin 3 Field,''
  %\href{http://www.slac.stanford.edu/spires/find/hep/www?irn=614335}{SPIRES entry}
  {\it  In *Singapore 1978, Proceedings, 1978 International Meeting On Frontier Of
  Physics, Vol.2*, Singapore 1978, 1101-1107}
  D.~P.~O'Brien,
  %``A Lagrangian Formalism For The Theory Of Spin 3 Fields,''
  Phys.\ Rev.\  D {\bf 18}, 4548 (1978).
  %%CITATION = PHRVA,D18,4548;%%
  M.~Kawasaki and M.~Kobayashi,
  %``Lagrange Formulation Of The Symmetric Theory Of Massive Spin 3 Fields. 1,''
  Phys.\ Rev.\  D {\bf 17}, 446 (1978).
  %%CITATION = PHRVA,D17,446;%%
\bibitem{gaugespin3}
  T.~Damour and S.~Deser,
  %``'GEOMETRY' OF SPIN 3 GAUGE THEORIES,''
  Annales Poincare Phys.\ Theor.\  {\bf 47}, 277 (1987).
  %%CITATION = AHPAA,47,277;%%
  A.~K.~H.~Bengtsson,
  %``On Gauge Invariance For Spin 3 Fields,''
  Phys.\ Rev.\  D {\bf 32}, 2031 (1985).
  %%CITATION = PHRVA,D32,2031;%%
\bibitem{emspin3}
  S.~M.~Klishevich,
  %``Electromagnetic interaction of massive spin-3 state from string theory,''
  Int.\ J.\ Mod.\ Phys.\  A {\bf 15}, 395 (2000)
  [arXiv:hep-th/9805174].
  %%CITATION = IMPAE,A15,395;%%
\bibitem{grspin3}
  Yu.~M.~Zinoviev,
  %``On spin 3 interacting with gravity,''
  arXiv:0805.2226 [hep-th].
  %%CITATION = ARXIV:0805.2226;%%
\bibitem{js}
  K.~Johnson and E.~C.~G.~Sudarshan,
  %``Inconsistency of the local field theory of charged spin 3/2 particles,''
  Annals Phys.\  {\bf 13}, 126 (1961).
  %%CITATION = APNYA,13,126;%%
\bibitem{spm3half}
  M.~Seetharaman, J.~Prabhakaran and P.~M.~Mathews,
  %``Rarita-Schwinger Particles In Homogeneous Magnetic Fields, And
  %Inconsistencies Of Spin 3/2 Theories,''
  Phys.\ Rev.\  D {\bf 12}, 458 (1975);
  %%CITATION = PHRVA,D12,458;%%
  %``Causality And Indefiniteness Of Charge In Spin 3/2 Field Theories,''
  J.\ Phys.\ A  {\bf 8}, 560 (1975).
  %%CITATION = JPAGB,A8,560;%%
\bibitem{rs3half}
  S.~D.~Rindani and M.~Sivakumar,
  %``Consistent Theory Of Massive Spin 3/2 Particle With Electromagnetic And
  %Gravitational Interaction By Kaluza-Klein Reduction,''
  J.\ Phys.\ G {\bf 12}, 1335 (1986).
  %%CITATION = JPHGB,G12,1335;%%
  %``KALUZA-KLEIN REDUCTION AND CONSISTENCY OF THE MASSIVE SPIN 3/2 THEORY WITH
  %EXTERNAL INTERACTION,''
  Z.\ Phys.\  C {\bf 49}, 601 (1991);
  %%CITATION = ZEPYA,C49,601;%%
\bibitem{d3}
  S.~Deser, V.~Pascalutsa and A.~Waldron,
  %``Massive spin 3/2 electrodynamics,''
  Phys.\ Rev.\  D {\bf 62}, 105031 (2000)
  [arXiv:hep-th/0003011].
  %%CITATION = PHRVA,D62,105031;%%
\bibitem{em3half}
  K.~Dormuth and Y.~Takahashi,
  %``Electromagnetic interaction of a spin 3/2 field,''
  Prog.\ Theor.\ Phys.\  {\bf 44}, 1077 (1970).
  %%CITATION = PTPKA,44,1077;%%
  A.~Z.~Capri and A.~Shamaly,
  %``Electrodynamics and propagation of a spin 3/2 field,''
  Can.\ J.\ Phys.\  {\bf 52}, 919 (1974).
  %%CITATION = CJPHA,52,919;%%
  R.~M.~Doria, J.~A.~Helayel-Neto and S.~Mokhtari,
  %``The Minimal coupling of charged spin 3/2 fields to an extended Abelian
  %gauge model,''
  Commun.\ Theor.\ Phys.\  {\bf 21}, 121 (1994).
  %%CITATION = CTPMD,21,121;%%
\bibitem{kli2}
  S.~M.~Klishevich,
  %``Massive fields of arbitrary half-integer spin in constant  electromagnetic
  %field,''
  Int.\ J.\ Mod.\ Phys.\  A {\bf 15}, 609 (2000)
  [arXiv:hep-th/9811030].
  %%CITATION = IMPAE,A15,609;%%
\bibitem{n2}
  P.~Van Nieuwenhuizen,
  %``Supergravity,''
  Phys.\ Rept.\  {\bf 68}, 189 (1981).
  %%CITATION = PRPLC,68,189;%%
\bibitem{n3}
  F.~A.~Berends, J.~W.~van Holten, P.~van Nieuwenhuizen and B.~de Wit,
  %``On Field Theory For Massive And Massless Spin 5/2 Particles,''
  Nucl.\ Phys.\  B {\bf 154}, 261 (1979);
  %%CITATION = NUPHA,B154,261;%%
  %``On Spin 5/2 Gauge Fields,''
  Phys.\ Lett.\  B {\bf 83}, 188 (1979)
  [Erratum-ibid.\  {\bf 84B}, 529 (1979)].
  %%CITATION = PHLTA,B83,188;%%
\bibitem{p5half}
  M.~Porrati,
  %``Massive spin 5/2 fields coupled to gravity: Tree level unitarity versus the
  %equivalence principle,''
  Phys.\ Lett.\  B {\bf 304}, 77 (1993)
  [arXiv:gr-qc/9301012].
  %%CITATION = PHLTA,B304,77;%%
\bibitem{gr5half}
  K.~Shima,
  %``On Gravitational Interaction Of Massless Spin 5/2 Particle,''
  Phys.\ Lett.\  B {\bf 124}, 321 (1983).
  %%CITATION = PHLTA,B124,321;%%
  R.~R.~Metsaev,
  %``Gravitational and higher-derivative interactions of massive spin 5/2 field
  %in (A)dS space,''
  Phys.\ Rev.\  D {\bf 77}, 025032 (2008)
  [arXiv:hep-th/0612279].
  %%CITATION = PHRVA,D77,025032;%%
\bibitem{w64}
  S.~Weinberg,
  %``Photons And Gravitons In S Matrix Theory: Derivation
%Of Charge Conservation
  %And Equality Of Gravitational And Inertial Mass,''
  Phys.\ Rev.\  {\bf 135}, B1049 (1964).
  %%CITATION = PHRVA,135,B1049;%%
\bibitem{zinolast}
  Yu.~M.~Zinoviev,
  %``On massive spin 2 electromagnetic interactions,''
  arXiv:0901.3462 [hep-th].
  %%CITATION = ARXIV:0901.3462;%%
\bibitem{aahdnr}
  A.~Adams, N.~Arkani-Hamed, S.~Dubovsky, A.~Nicolis and R.~Rattazzi,
  %``Causality, analyticity and an IR obstruction to UV completion,''
  JHEP {\bf 0610}, 014 (2006)
  [arXiv:hep-th/0602178].
  %%CITATION = JHEPA,0610,014;%%
\bibitem{hm}
  D.~M.~Hofman and J.~Maldacena,
  %``Conformal collider physics: Energy and charge correlations,''
  JHEP {\bf 0805}, 012 (2008)
  [arXiv:0803.1467 [hep-th]].
  %%CITATION = JHEPA,0805,012;%%
\bibitem{dix}
J.~A.~Dixon,
  %``Calculation of BRS cohomology with spectral sequences,''
  Commun.\ Math.\ Phys.\  {\bf 139}, 495 (1991).
  %%CITATION = CMPHA,139,495;%%
\end{thebibliography}
\end{document}